\documentclass[useAMS,usenatbib,pdfsync]{mn2e}

\usepackage{amstext}
\usepackage{amsmath, float}
\usepackage{graphicx}

\newcommand       \be           {\begin{equation}}
\newcommand       \ee           {\end{equation}}
\newcommand       \bea          {\begin{eqnarray}}
\newcommand       \eea          {\end{eqnarray}}

\def\simlt{\mathrel{\hbox{\rlap{\hbox{\lower4pt\hbox{$\sim$}}}\hbox{$<$}}}}
\def\simgt{\mathrel{\hbox{\rlap{\hbox{\lower4pt\hbox{$\sim$}}}\hbox{$>$}}}}
\def\simlt{\mathrel{\hbox{\rlap{\hbox{\lower4pt\hbox{$\sim$}}}\hbox{$<$}}}}
\def\simgt{\mathrel{\hbox{\rlap{\hbox{\lower4pt\hbox{$\sim$}}}\hbox{$>$}}}}
\def\simlt{\mathrel{\hbox{\rlap{\hbox{\lower4pt\hbox{$\sim$}}}\hbox{$<$}}}}
\def\simgt{\mathrel{\hbox{\rlap{\hbox{\lower4pt\hbox{$\sim$}}}\hbox{$>$}}}}
\def\lesssim{\mathrel{\hbox{\rlap{\hbox{\lower4pt\hbox{$\sim$}}}\hbox{$<$}}}}
\def\gtrsim{\mathrel{\hbox{\rlap{\hbox{\lower4pt\hbox{$\sim$}}}\hbox{$>$}}}}

\def\simlt{\mathrel{\hbox{\rlap{\hbox{\lower4pt\hbox{$\sim$}}}\hbox{$<$}}}}
\def\simgt{\mathrel{\hbox{\rlap{\hbox{\lower4pt\hbox{$\sim$}}}\hbox{$>$}}}}

\title[SGRBEEs from proto-magnetar spin-down]{Short GRBs with Extended Emission from Magnetar Birth: Jet Formation and Collimation}
\author[N. Bucciantini, et al.]{N. Bucciantini$^{1}$\thanks{E-mail:niccolo@nordita.org}, B.~D. Metzger$^{2,5}$,
 T.~A. Thompson$^{3,6}$, E. Quataert$^{4}$\\
  $^{1}$NORDITA, AlbaNova University Center, Roslagstullsbacken 23, SE 10691 Stockholm, Sweden\\ 
$^{2}$Department of Astrophysical Sciences, Peyton Hall, Princeton
University, Princeton, NJ 08544, USA\\ 
$^{3}$Department of Astronomy and Center for Cosmology \&
Astro-Particle Physics, The Ohio State University, Columbus, OH 43210,
USA\\ 
$^{4}$ Astronomy Department and Theoretical Astrophysics Center,
  University of California, Berkeley, 601 Campbell Hall,\\ Berkeley CA,
  94720\\
$^{5}$ NASA Einstein Fellow\\
$^{6}$ Alfred P. Sloan Fellow}
\begin{document}

\date{Accepted . Received ; in original form }

\pagerange{\pageref{firstpage}--\pageref{lastpage}} \pubyear{????}

\maketitle

\label{firstpage}

\begin{abstract}

Approximately $1/4-1/2$ of short duration Gamma-Ray Bursts (GRBs) are followed by variable X-ray emission lasting $\sim 100$ s with a fluence comparable or exceeding that of the initial burst itself.  The long duration and significant energy of this `extended emission' (EE) poses a major challenge to the standard binary neutron star (NS) merger model.  \citet{metzger08} recently proposed that the EE is powered by the spin-down of a strongly magnetized neutron star (a {\it millisecond proto-magnetar}), which either survives the NS-NS merger or is created by the accretion-induced collapse of a white dwarf.  However, the effects of surrounding material on the magnetar outflow have not yet been considered.  Here we present time-dependent axisymmetric relativistic magnetohydrodynamic simulations of the interaction of the relativistic proto-magnetar wind with a surrounding $10^{-1}-10^{-3} M_\odot$ envelope, which represents material ejected during the merger; in the supernova following AIC; or via outflows from the initial accretion disk.  The collision between the relativistic magnetar wind and the expanding ejecta produces a termination shock and a magnetized nebula inside the ejecta.  A strong toroidal magnetic field builds up in the nebula, which drives a bipolar jet out through the ejecta, similar to the magnetar model developed in the case of long duration GRBs.  We quantify the `break-out' time and opening angle of the jet $\theta_{\rm j}$ as a function of the wind energy flux $\dot{E}$ and ejecta mass $M_{\rm ej}$.  We show that $\dot{E}$ and $\theta_{\rm j}$ are inversely correlated, such that the beaming-corrected (isotropic) luminosity of the jet (and hence the observed EE) is primarily a function of $M_{\rm ej}$.  Both variability arguments, and the lower limit on the power of magnetar outflows capable of producing bright emission, suggest that the true opening angle of the magnetar jet must be relatively large.  The model thus predicts a class of events for which the EE is observable with no associated short GRB.  These may appear as long-duration GRBs or X-Ray Flashes unaccompanied by a bright supernova and not solely associated with massive star formation, which may be detected by future all-sky X-ray survey missions.

\end{abstract}


\section{Introduction}
\label{sec:intro}

Gamma-Ray Bursts (GRBs) are canonically divided into `long soft' (LGRB) and `short hard' (SGRB) classes, based on their bimodal distribution in duration and spectral hardness \citep{kouveliotou93}.  This division is supported by studies of the host galaxies and environments of GRBs.  LGRBs occur in late type galaxies with high specific star formation rates (e.g. \citealt{Fruchter+06}; \citealt{Levesque+10}) and are accompanied by core collapse supernovae (SNe) \citep{galama98,dellavalle06,chornock10,starling10}.  In contrast, SGRBs occur in both elliptical (\citealt{bloom06}; \citealt{Berger+05}) and late type (\citealt{Fox05}; \citealt{Barthelmy+05}) galaxies; they show larger average offsets from their host centres (e.g.~\citealt{Prochaska+06}; \citealt{berger09,fong10}); and no evidence is found for a bright associated supernova in a few well-studied cases (e.g.~\citealt{Hjorth+05}; \citealt{Kann+08}). 

The differences between long and short GRBs have motivated the development of distinct, albeit related, models for their central engines.  LGRBs originate when the core of a rotating massive star collapses to form either a black hole (BH) \citep{macfadyen99} or a rapidly spinning, strongly magnetized neutron star (a {\it millisecond proto-magnetar}; e.g.~\citealt{Usov92}; \citealt{wheeler00}; \citealt{Thompson+04}).  SGRBs are instead commonly attributed to the inspiral and merger of neutron star (NS)-NS or NS-BH binaries (\citealt{lee07}; \citealt{Nakar07}), although the accretion induced collapse (AIC) of a White Dwarf (WD) \citep{metzger08} or a NS \citep{dermer06} represent viable alternatives.  In all SGRB models, the short $\sim 0.1-1$ s duration of the burst is related to the accretion timescale of the compact $\sim 10^{-3}-0.1M_{\sun}$ torus that forms around the central NS or BH following the merger (e.g.~\citealt{Janka+99}) or AIC event (\citealt{Dessart+06}).

The standard LGRB/SGRB dichotomy has recently been challenged by several `hybrid' events that conform to neither class (e.g.~\citealt{zhang07}; \citealt{Bloom+08}).  GRB 060505 and 060614 are both long bursts based on their duration, yet neither shows evidence for a bright associated SN (\citealt{fynbo06}; \citealt{Gehrels+06}; \citealt{Gal-Yam+06}; \citealt{Ofek+07}).  Although the short GRB 050724 occurred in an elliptical host with no associated SN, it was followed by variable X-ray emission lasting $\sim 140$ s with a total fluence $\sim 3$ times greater than the short GRB itself \citep{Barthelmy+05}.  Similar to 060614, the light curve of GRB 080503 was characterized by a hard initial spike, followed (after a brief lull) by a `hump' of X-ray emission lasting $\sim 100$ s and carrying $\sim 30$ times the fluence of the initial spike \citep{perley09}.  The rapid variability of this `extended emission' (EE) strongly suggests that it results from ongoing central engine activity.  

All together approximately $\sim 1/4$ of {\it Swift} SGRBs\footnote{When observational bias due to the effects of e.g.~detection threshold are taken into account, the true fraction of SGRBEEs could be as high as $\sim 50\%$ \citep{Norris+10}.} are accompanied by extended X-ray emission lasting for $\sim 10-100$ s with a fluence $\gtrsim$ that of the GRB itself (see \citealt{norris08} and \citealt{perley09} for a compilation of events).   The hybrid nature and common properties of these events (`short GRB' + $\sim$ 100 s X-ray tail) have motivated the introduction of a new subclass: Short GRBs with Extended Emission (SGRBEEs).  It was moreover recently discovered that some SGRBs are followed by an X-ray `plateau' ending in a very sharp break (GRB 980515; \citealt{rowlinson10}). \citealt{Troja+08}; \citealt{Lyons+10}) and is difficult to explain by circumstellar interaction alone.  Although the connection of this event to SGRBEEs is unclear, it nevertheless provides additional evidence that the central engine is active at late times.

An important observational question is whether SGRBEEs differ from `normal' short bursts in other properties.  \citet{Troja+08} found that SGRBEEs occur on average closer to the centres of their host galaxies than other SGRBs; however, the current paucity of well-localized events make statistical claims uncertain and recent studies have not verified this result \citep{berger09,fong10}.  \citet{norris11} showed that SGRBEEs differ also in the properties of the the initial short GRB itself.  They find that the average duration of the burst, and of individual pulse structures, are longer for short GRBs with EEs, possibly suggesting the existence of different progenitor channels (\citealt{Leibler&Berger10}; \citealt{virgili11}) or a different circumburst environment.

The long duration and high fluence of the extended emission of SGRBEEs poses a serious challenge to the NS merger scenario, because in this model both the prompt and extended emission are necessarily powered by black hole accretion.  It is in particular difficult to understand how such a high accretion rate is maintained at very late times.  Although accretion of the torus formed from the merger may power the short GRB itself (e.g.~\citealt{Janka+99}), the timescale for the disk to accrete is short (typically $\lesssim 1$ s), and the disk is disrupted by outflows soon thereafter (\citealt*{Metzger+09a}; \citealt*{Metzger+09b}; \citealt{Lee+09}).  `Fall back' accretion could in principle power late emission (\citealt{Rosswog07}; \citealt{Faber+06}; \citealt{Chawla+10}), but whether sufficient mass is placed onto the highly eccentric (yet bound) orbits required by this scenario is unclear.  Full simulations of the fall-back process, including the important effect of heating due to $r$-process nucleosynthesis \citep*{Metzger+10a}, have yet to be performed.  Even if sufficient mass returns at late times, the accretion will occur under radiatively inefficient conditions (e.g.~\citealt{Narayan+01}); since the disk is only marginally bound when it cannot cool, most of the accreting mass may again be lost to [non-relativistic] outflows \citep{Rossi&Begelman09}.

\begin{figure}
\resizebox{\hsize}{!}{\includegraphics{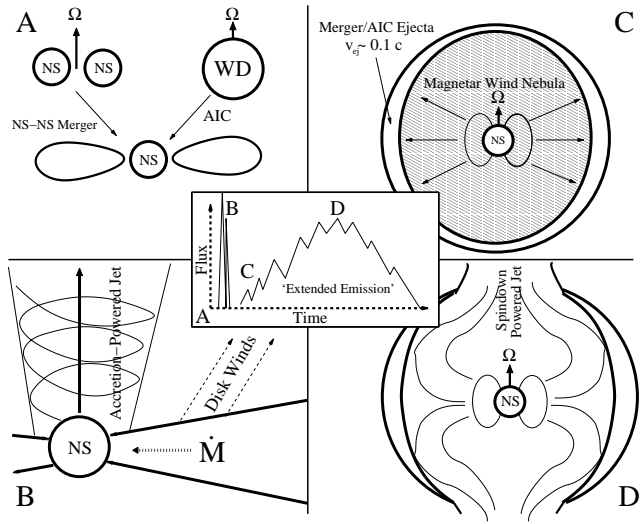}}
\caption{Schematic diagram of the stages of the proto-magnetar model for Short GRBs with Extended Emission.  (A) The merger of two binary neutron stars, or the accretion-induced collapse of a rotating white dwarf, results in the formation of a compact $\sim 10^{-3}-0.1M_{\sun}$ torus around the central proto-neutron star. (B) Accretion of the torus powers a relativistic bipolar jet, resulting in a short GRB lasting $\sim 0.1-1$ s, similar to the standard NS-NS merger model.  Following accretion, however, a rapidly spinning (millisecond) {\it proto-magnetar} remains.  (C)  Material ejected during the merger, by the supernova following AIC, or via outflows from the accretion disk, results in a $\sim 10^{-3}-10^{-1}M_{\sun}$ envelope around the proto-magnetar moving outwards with a velocity $v_{\rm ej} \sim 0.1-0.2$ c.  The relativistic wind from the proto-magnetar collides with the ejecta, producing a {\it magnetar wind nebula}.  (D) Magnetic stresses in the nebula redirect the magnetar wind into a bipolar jet.  After the jet breaks through the ejecta on a timescale $\sim 1-10$ s (Fig.~\ref{fig:and}), the magnetar wind escapes and accelerates to ultrarelativistic speeds (Fig.~\ref{fig:jets}).  Emission from the jet at much larger radii powers the extended emission lasting $\sim 10-100$ s, similar to the proto-magnetar model for long GRBS (see Fig.~\ref{fig:lcs}).}
\label{fig:cartoon}
\end{figure}

\subsection{The Proto-Magnetar Model for SGRBEEs}

The hybrid properties of SGRBEEs hint that the engine could itself be of hybrid nature.\footnote{\citet{lazzati10} recently proposed that SGRBEEs may result from collapsars viewed off-axis.  However, the differences between the host galaxies and environments between SGBEEs and long GRBs \citep{berger09,fong10,Leibler&Berger10}, and the lack in some events of an associated supernovae, indicates that this is not the main channel.}  \citet{metzger08} recently proposed that SGRBEEs result from the birth of a rapidly spinning proto-magnetar, created by a NS-NS merger or the AIC of a WD (see Figure \ref{fig:cartoon} for an schematic illustration).  In this model the short GRB is powered by the accretion of the initial torus (similar to standard NS merger models), but the EE is powered by a relativistic wind from the proto-magnetar at later times, after the disk is disrupted.  Although a NS remnant is guaranteed in the case of AIC, the merger of a double NS binary could also leave a stable NS remnant, provided that either (1) the total mass of the binary is low and/or the NS equation of state is stiff \citep{Shibata&Taniguchi06}; (2) the proto-NS forms in a meta-stable state supported by differential rotation \citep{Baumgarte+00,baiotti+08}, but it then loses sufficient mass via magneto-centrifugal outflows (\citealt{Thompson+04}; \citealt*{Metzger+07}) to reach stability.  The likelihood of this possibility has increased recently due to the discovery of a $\approx 2M_{\sun}$ NS (\citealt{Demorest+10}), which suggests that the nuclear EOS is indeed stiff (see also \citealt{Ozel+10}).  Given that rapid rotation is expected in both NS-NS merger and AIC scenarios, it is plausible that the proto-NS will generate a magnetar-strength field by, for instance, an $\alpha-\Omega$ dynamo \citep{Duncan&Thompson92}, shear instabilities at the merger interface \citep{Price&Rosswog06}, or the magneto-rotational instability (MRI; e.g.~\citealt{Akiyama+03}; \citealt{Thompson+05}).

Though not surrounded by the envelope of a massive star, magnetars formed from NS-NS mergers or AIC do not form in vacuum.  In the AIC case $\sim 10^{-3}-10^{-2}M_{\sun}$ is ejected during the SN explosion on a timescale $\lesssim 1$ s (e.g.~\citealt{Woosley&Baron92}; \citealt{Dessart+06}), while in NS-NS mergers a similar mass may be ejected dynamically due to tidal forces during the merger process (e.g.~\citealt{Rosswog07}).  Mass loss also occurs in outflows from the accretion disk on timescales $\lesssim$ seconds, due to heating from neutrinos (\citealt*{Metzger+08a}; \citealt{Dessart+09}), turbulent viscosity (\citealt*{Metzger+08b}; \citealt*{Metzger+09a}), and nuclear energy released by the recombination of free nuclei into $^{4}$He (\citealt{lee07}; \citealt*{Metzger+08b}; \citealt{Lee+09}).  During the first few seconds after forming, outflows from the magnetar itself are heavily mass-loaded and non-relativistic, resulting in a significant quantity of ejecta $\gtrsim 10^{-3}M_{\odot}$ \citep{Thompson+04,Bucciantini+06,Metzger+07}.  All together, $\sim 10^{-3}-0.1M_{\sun}$ is ejected with a characteristic velocity $v_{\rm ej} \sim 0.1-0.2$ c and kinetic energy $\sim 2\times 10^{50}(v_{\rm ej}/0.1c)^{2}(M_{\rm ej}/0.01M_{\sun})$ ergs. 

A few seconds after the merger or AIC, one is left with a proto-magnetar embedded in a confining envelope.\footnote{In cases when the ejecta originates from the earlier [non-relativistic] stage of the magnetar wind, the distinction between `wind' and `ejecta' is blurred.  In general, however, the magnetar outflow becomes ultra-relativistic relatively abruptly, such that this distinction is well-defined \citep{metzger10}.}  This configuration is qualitatively similar to that developed in the proto-magnetar model for LGRBs by \citet{bucciantini07,bucciantini08,bucciantini09}, except that the enshrouding envelope is much less massive.  In these previous works it was shown that, although the power in the magnetar wind is relatively isotropic (e.g.~\citealt{Bucciantini+06}), its collision with the slowly-expanding ejecta produces a hot `proto-magnetar nebula' \citep{bucciantini07}.  As toroidal flux accumulates in the nebula, magnetic forces -- and the anisotropic thermal pressure they induce -- redirect the equatorial outflow towards the poles (\citealt{Begelman&Li92}; \citealt{Konigl&Granot02}; \citealt{Uzdensky&MacFadyen07}; \citealt{bucciantini07,bucciantini08,bucciantini09}; \citealt{Komissarov&Barkov07}).  Stellar confinement thus produces a mildly-relativistic jet, which drills a bipolar cavity through the ejecta.  Once the jet `breaks out', an ultra-relativistic jet (fed by the magnetar wind at small radii) freely escapes.  The EE is then powered as the jet dissipates its energy at much larger radii.  One virtue of applying this picture to SGRBEEs is that it naturally explains why the EE resembles long GRBs in several properties, such as its duration and the existence of a late-time `steep decay' phase (cf.~\citealt{Tagliaferri+05}; \citealt{perley09}).  

Although SGRBEEs resemble long GRBs in many properties, important differences also exist.  The EE is generally softer (X-rays rather than gamma-rays), somewhat dimmer, and its variability is generally smoother (appearing to display e.g.~a higher `duty cycle'), than long GRBs.  Assessing the viability of the proto-magetar model for SGRBEEs therefore requires determining whether these differences may in part result from differences in the geometry of the relativistic outflow.  These in turn may result because the confining ejecta is significantly less massive and dense than in the core collapse case.

In this paper we investigate the interaction of the relativistic proto-magnetar wind with the expanding ejecta using axisymmetric (2D) relativistic MHD simulations.  We focus in particular on the confining role of the ejecta and its dependence on the wind power, and on the ejecta mass and density profile.  We show that collimation (jet formation) is achieved only within a bounded range of parameters.  If the wind is too energetic, or the mass of the shell is too low, the ejecta is disrupted and little collimation occurs.  In contrast, if the ejecta is sufficient massive and/or the wind is sufficient weak, the result is instead a `choked jet' that may not emerge at all.  We describe the numerical set-up in Section \ref{sec:num} and present our results in Section \ref{sec:results}.  We apply our results to SGRBEEs in Section \ref{sec:discussion} and conclude in Section \ref{sec:conclusions}.  

\section{Numerical Setup}
\label{sec:num}
All calculations were performed using the shock-capturing central-scheme for relativistic ideal MHD ECHO
\citep{delzanna03,delzanna07}, using an ideal gas equation of state with an adiabatic coefficient $\Gamma = 4/3$, as appropriate for relativistically hot gas.  We refer the reader to these papers for a detailed description of the equations and numerical algorithms.

We investigate the interaction of the magnetar wind with the surrounding ejecta envelope using 2D axisymmetric simulations on a spherical grid. The angular domain is $\theta = [0,\pi]$ with reflecting boundary at the polar axis to enforce axisymmetry, while the radial domain extends over the range $r=[10^7,10^{12}]$cm.  The grid in the radial direction is spaced logarithmically with 100 cells per decade, while spacing is uniform in the angular direction with 200 cells [we repeated selected simulations with twice the resolution to verify convergence; see also \citet{camus09} for estimates of convergence with grid resolution, in similar simulations as applied to pulsar wind nebulae (PWNe)].  We assume zeroth-order extrapolation at the outer boundary.  The code is second-order in both space and time, with a monotonized central limiter, chosen in order to resolve the large density jump between the lighter relativistic plasma inside the magnetar wind nebula (MWN) and the heavier envelope.

\citet{bucciantini07} showed that the interaction between a magnetar wind and a confining envelope depends on the strength of the toroidal magnetic field $B$ in the MWN; this in turn depends on the magnetization in the wind ($\sigma = r^2B^2 c/ \dot{E}$, where $\dot{E}$ is the wind energy flux) at the distance of the termination shock.  While $\sigma$ can be calculated at the light cylinder radius with some confidence (e.g.~Metzger et al.~2010), its value at larger radii is difficult to determine due to uncertainties in the conversion of magnetic energy into kinetic energy in relativistic winds (see \citealt{bucciantini07} for a detailed discussion of this problem in the context of LGRBs).  Moreover, instabilities may occur inside the MWN \citep{begelman98} which further reduce the toroidal magnetic field strength.  Nevertheless, previous studies \citep{bucciantini08,bucciantini09} show that reliable results for the dynamics of the MWN and the properties of the jet are obtained using even the simplified regime of transverse relativistic MHD.

For proto-magnetars with millisecond rotation periods, the light cylinder is located at $\sim 10^7$ cm and the fast magnetosonic surface is at $\sim 10^7-10^8$ cm \citep{Bucciantini+06}.  As in the calculations of \citet{bucciantini08}, at the inner boundary we inject a super-magnetosonic wind with a fixed Lorentz factor $\gamma = 25$ and a magnetization of $\sigma = 0.1$; these values are appropriate for distances of the order of the termination shock radius if the conversion of magnetic to kinetic energy is efficient, and for typical properties of the magnetar wind at a few seconds after the proto-neutron star forms (\citealt{metzger10}; see also the discussion in Sec.~\ref{sec:results} below).  Under these assumptions, $\sigma$ is conserved throughout the upstream region.  We assume that the wind contains a purely toroidal field and is cold with $\rho c^2/p= 100$, where $\rho$ and $p$ are the density and pressure, respectively.  For simplicity we assume that $\dot{E}$ and $\gamma$ are constant, and that the wind is isotropic, throughout the simulation.  Although in reality $\gamma$ increases from $\sim 1$ to $\sim 100-1000$ over tens of seconds as the proto-magnetar cools \citep{metzger10}, studies of PWNe show that the dynamics becomes independent of $\gamma$ in the limit $\gamma \gg 1$.

Following \citet{darbha10} we assume that the shell of ejecta expands radially homologously with a velocity $v_{\rm ej} =0.2 c(r/r_{\rm e})$ inside a low density cavity of radius $r_{\rm in}$, where $r_{\rm e}$ is the radius of the outer edge of the shell.  To this self-similar profile we further allow for a modulation with polar angle $\theta$:
\begin{equation}
\rho=\begin{cases} K(1-\alpha \cos{2\theta})r^{-C}& \text{if
    $r_{\rm in} < r
    < r_{\rm e}$},\\ 10^{-5}\text{g cm}^{-3}& \text{if $r<r_{\rm in}$}.
\label{eq:rho}
\end{cases}
\end{equation}
where $r_{\rm e}=6\times 10^9$cm, $r_{\rm in}=1.5\times 10^9$cm (corresponding to initial conditions one second after ejection), $C=4$, $K$ is fixed by the total ejecta mass $M_{\rm ej}$, and we have chosen the density in the inner cavity to be sufficiently low as to have a negligible effect on the dynamics.  The angular profile we adopt is equatorially concentrated, motivated by the possibilities that either (1) the ejecta is the result of an equatorially-focused disk wind (e.g.~\citealt*{Metzger+09a}); or (2) a bipolar asymmetry remains from the jet produced during the early accretion-powered (short GRB) phase.  Note that the self-similar profile of the ejecta implies that the ratio between the kinetic energy and mass of the ejecta is fixed at the value $E_{\rm ej}/M_{\rm ej}c^2\simeq 0.005$.  Outside $r_{\rm e}$ we assume a stationary, uniform medium with a density $\rho = 10^{-5}$g cm$^{-3}$, which we have verified is sufficiently tenuous that it has negligible effects on the dynamics and geometry of jet formation.

\section{Results}
\label{sec:results} 

\begin{figure*}
\resizebox{\hsize}{!}{\includegraphics{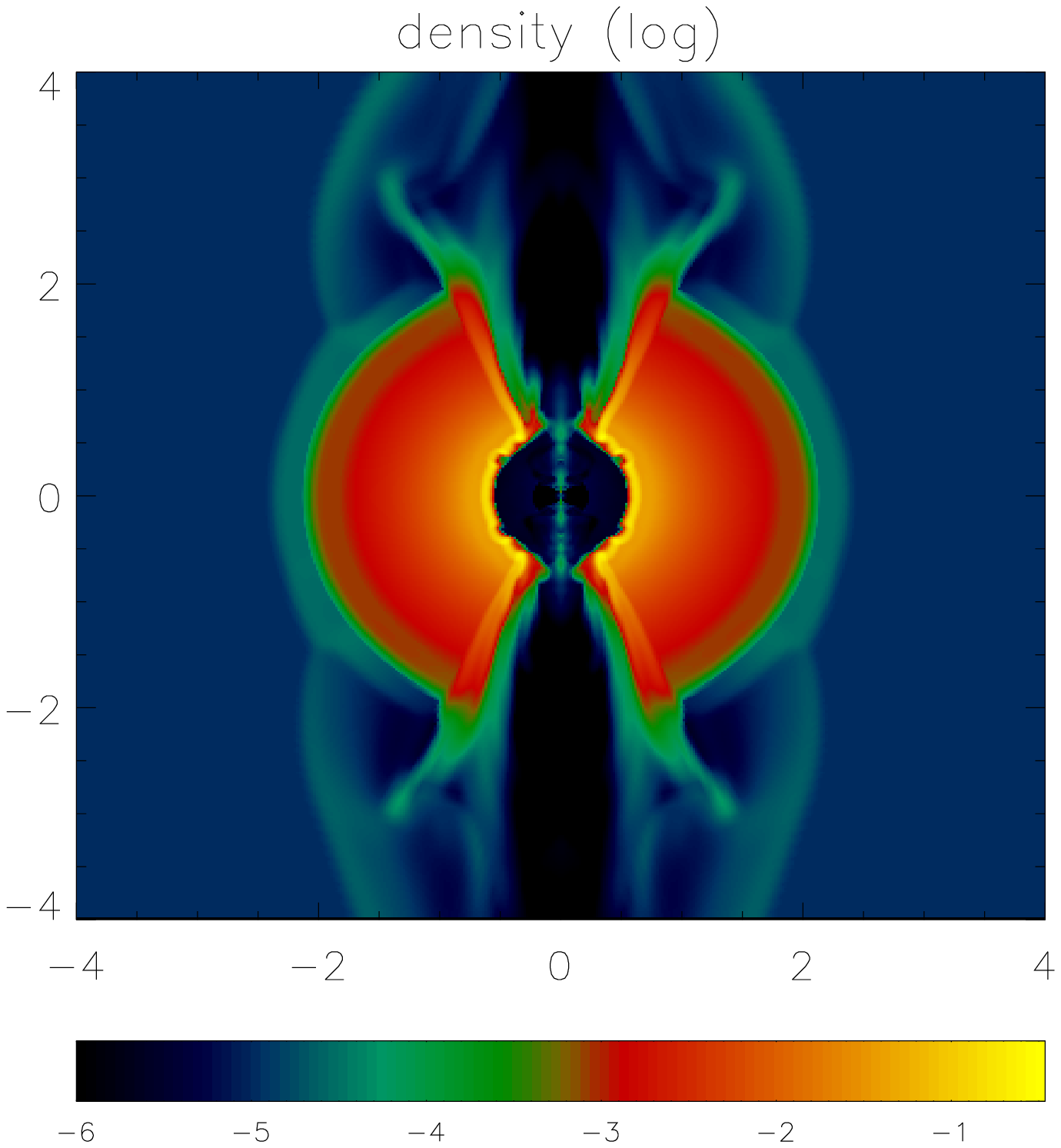}\includegraphics{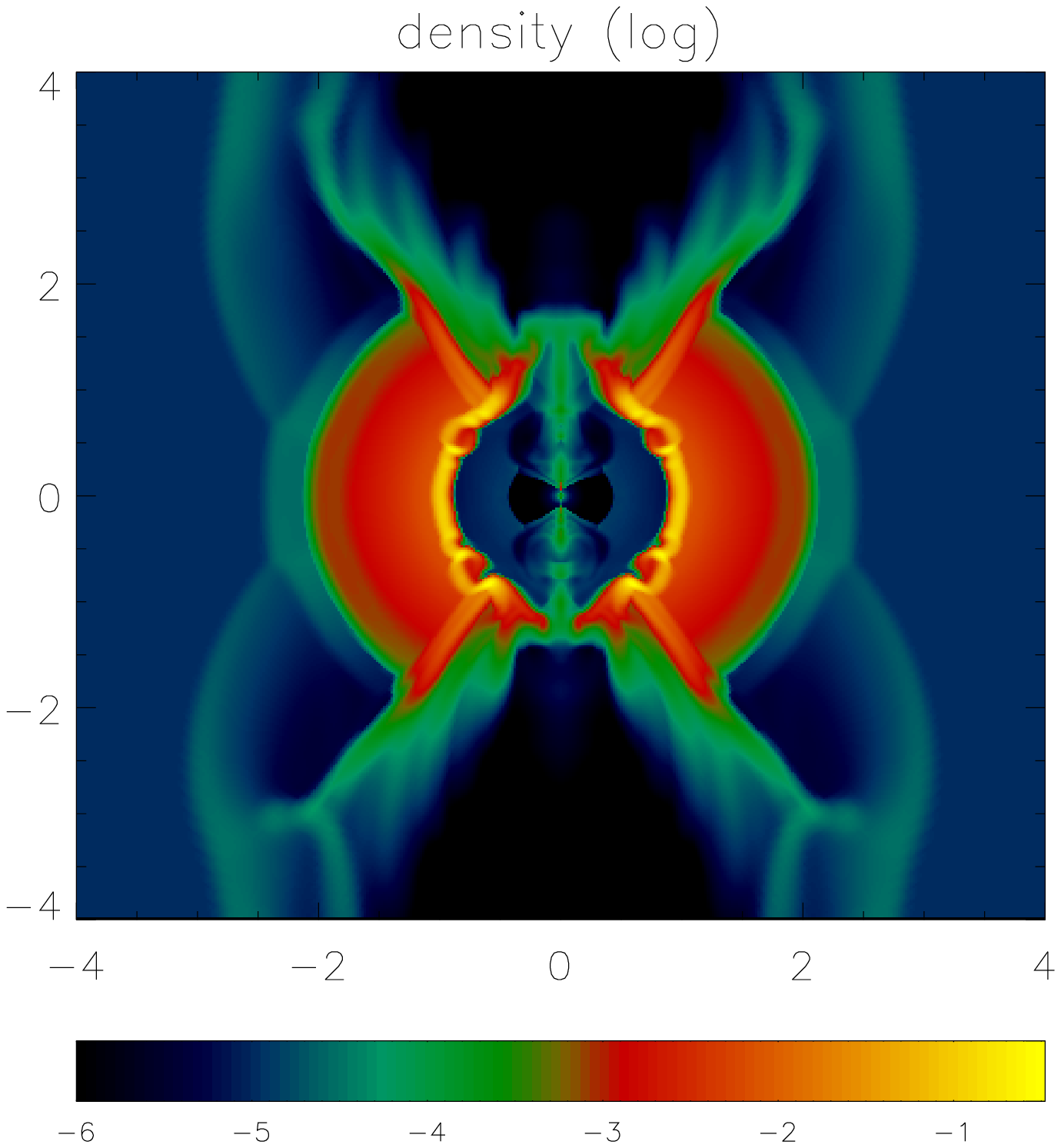}\includegraphics{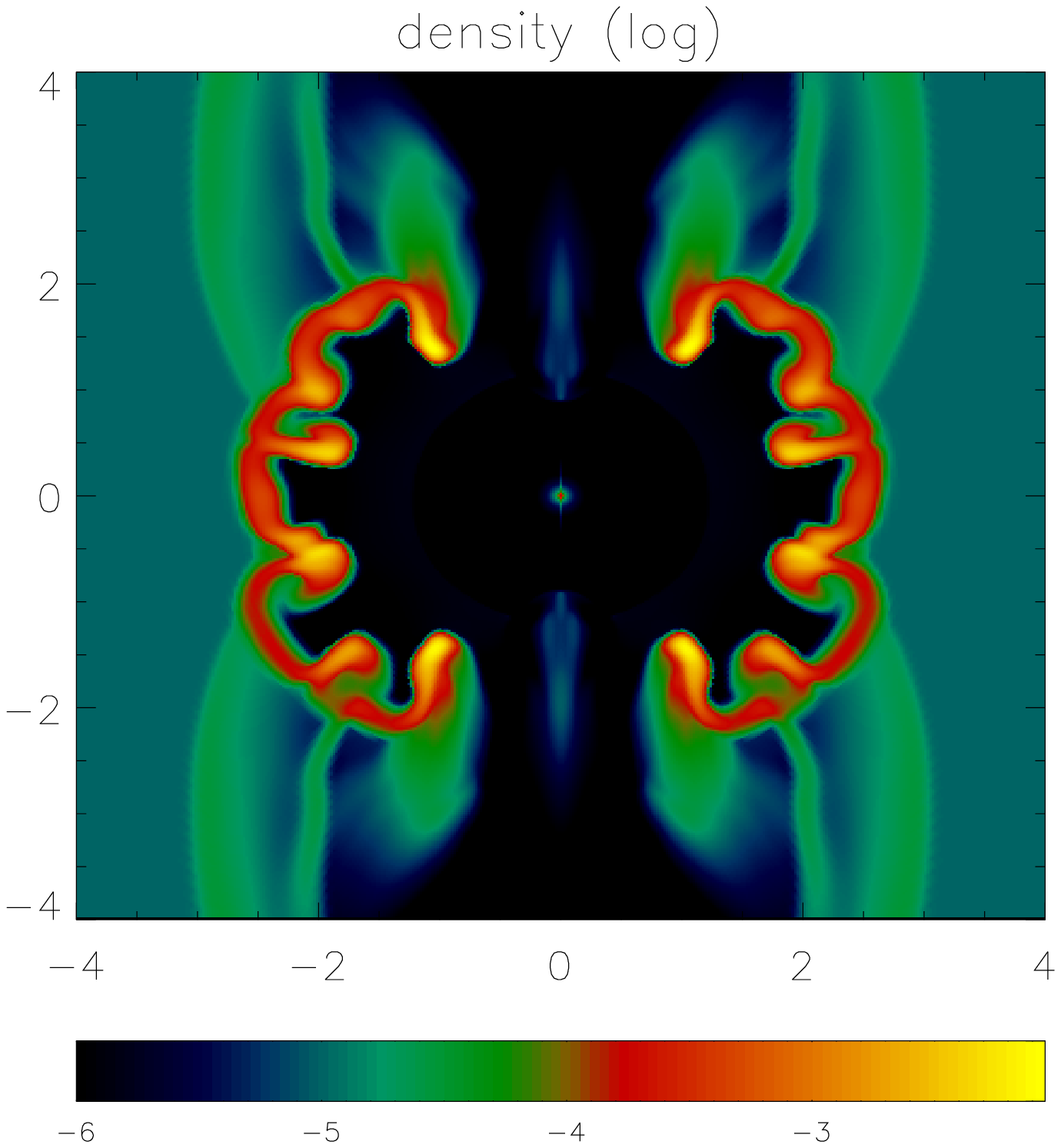}}
\resizebox{\hsize}{!}{\includegraphics{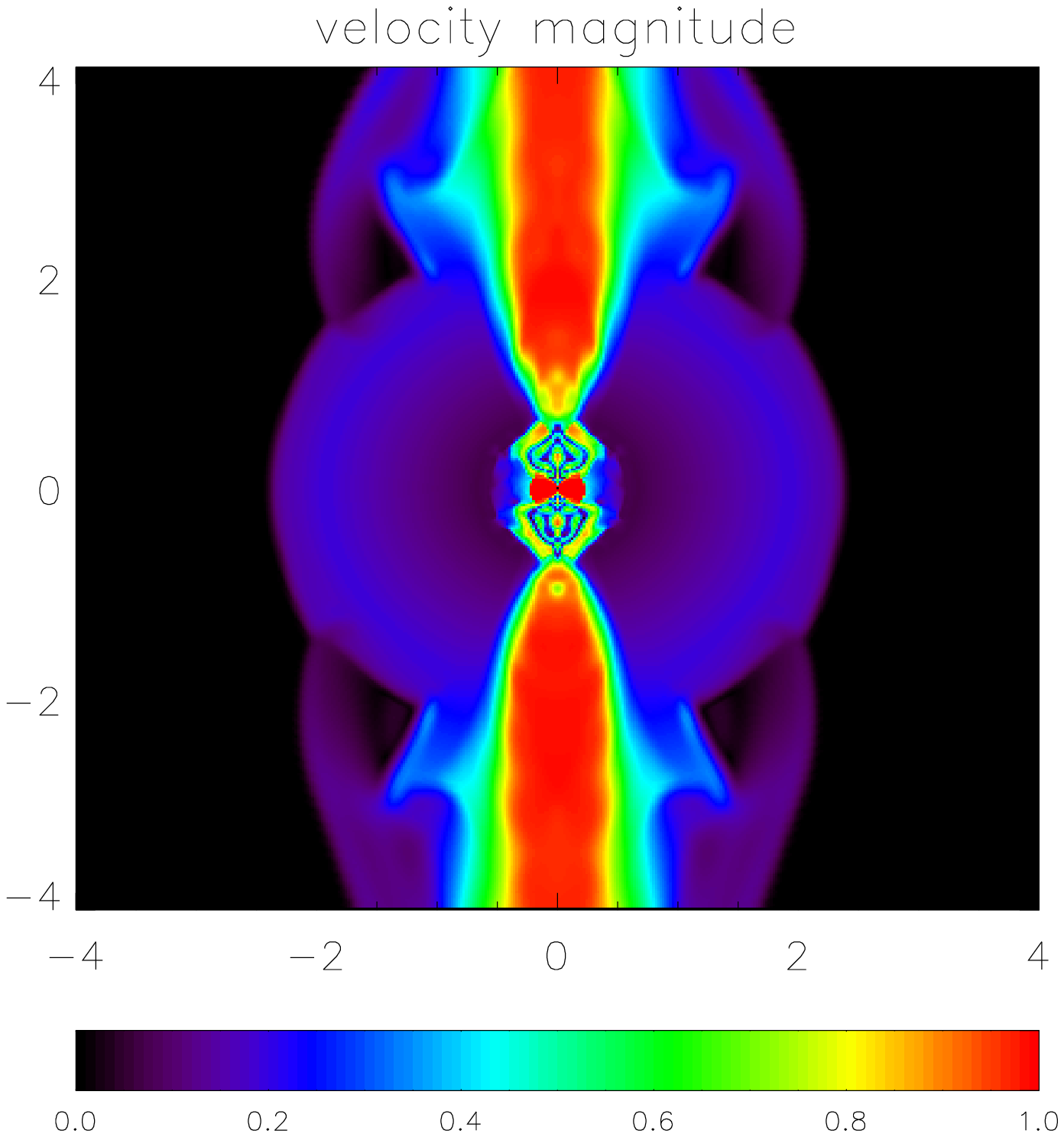}\includegraphics{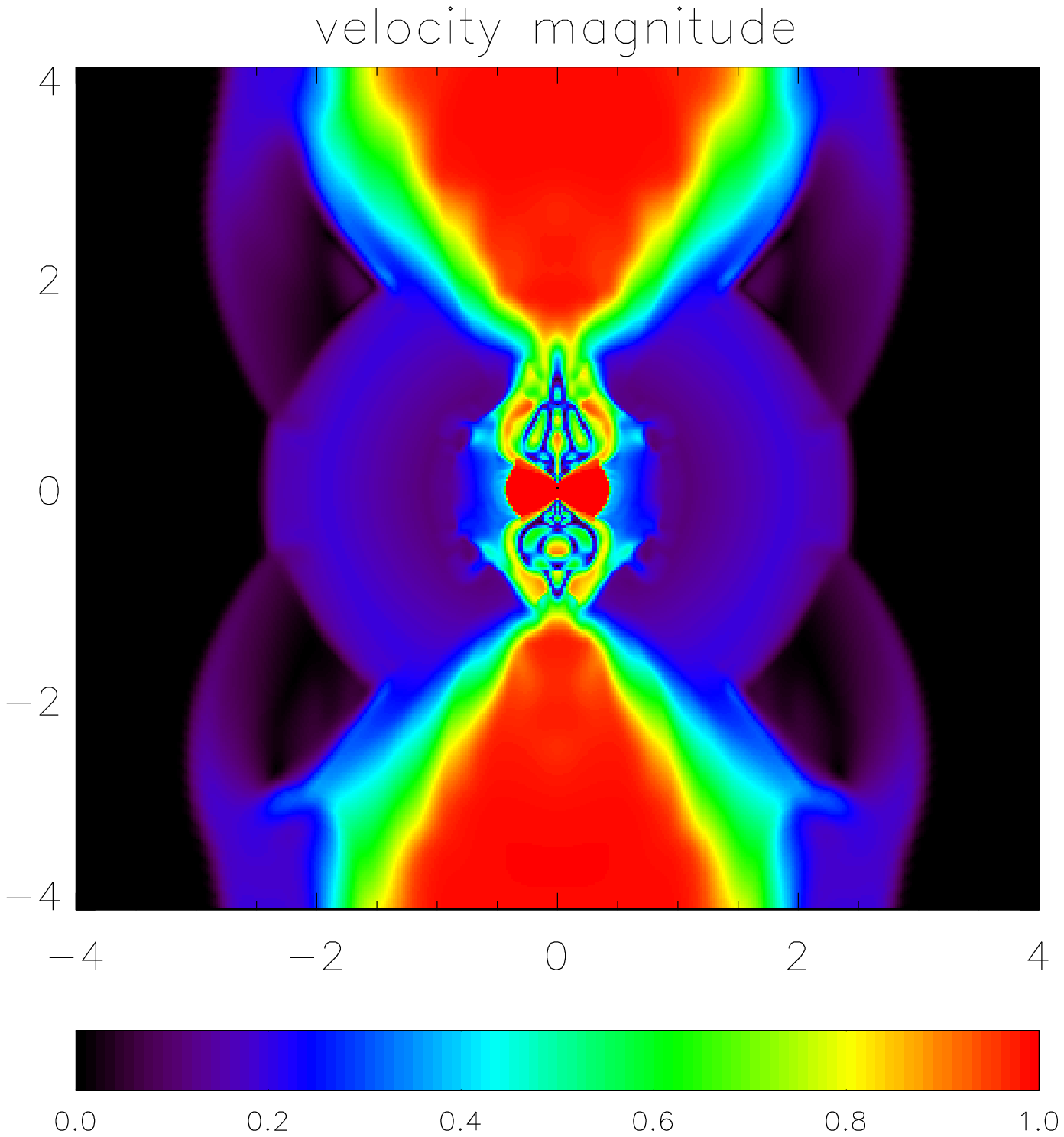}\includegraphics{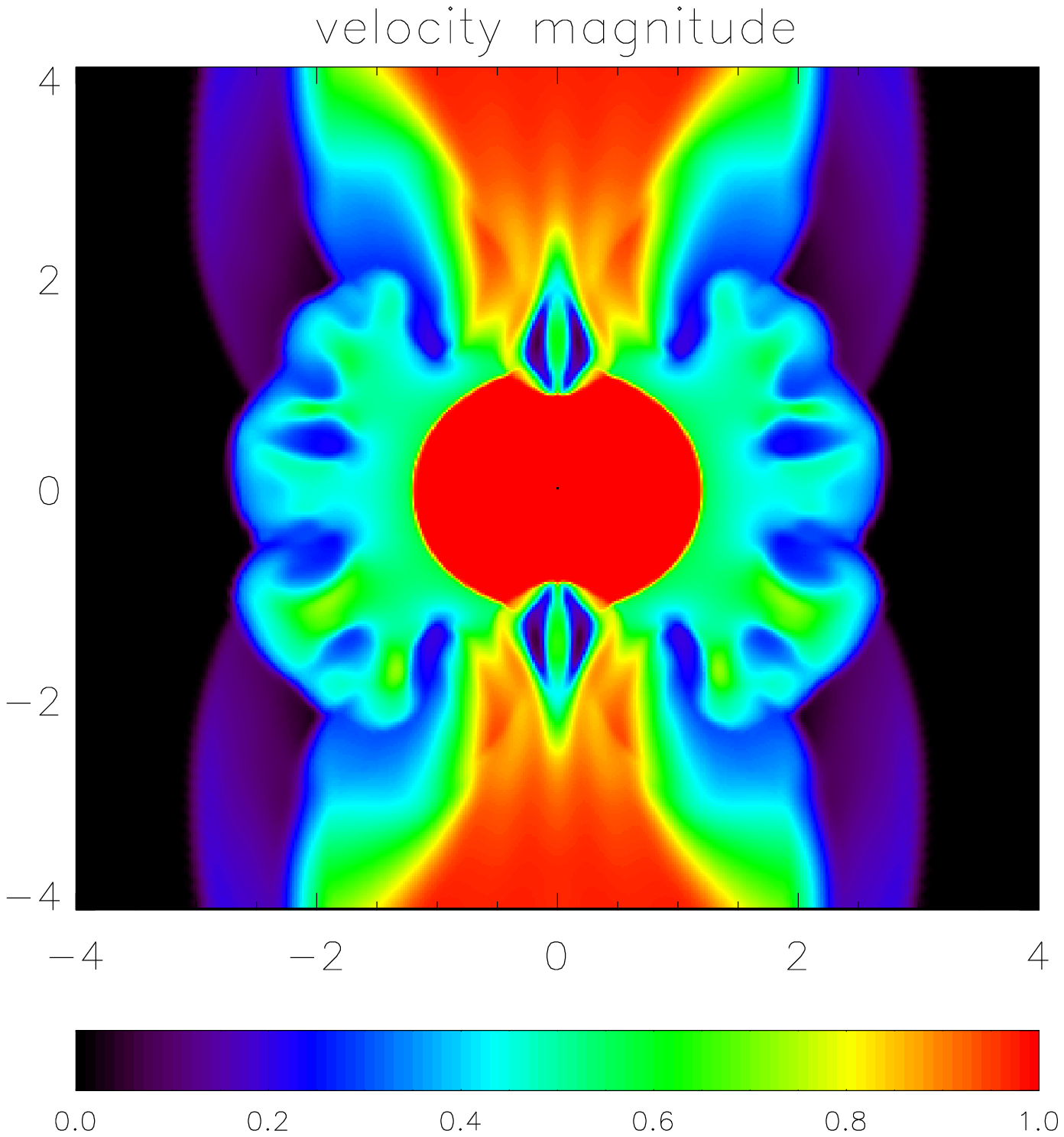}}
\caption{Snapshots of the interaction of the proto-magnetar wind with the confining shell of ejecta at $t=35$ seconds. The upper panel shows the density structure (cgs units), while the lower panel shows the magnitude of the velocity in units of the speed of light.  The left, centre, and right columns show, respectively, cases corresponding to a low ($\dot{E}/M_{\rm ej}=10^{50}$ erg s$^{-1}$ M$_\odot^{-1}$; case A), `average' ($\dot{E}/M_{\rm ej}=10^{51}$ erg s$^{-1}$ M$_\odot^{-1}$; case B) and high ($\dot{E}/M_{\rm ej}=10^{52}$ erg s$^{-1}$ M$_\odot^{-1}$; case C) power wind. Axis are in units of $10^{11}$ cm.}
\label{fig:jets}
\end{figure*}

Due to the self-similar nature of the ejecta, the dynamics of the wind-ejecta interaction is only a function of the ratio $\dot{E}/M_{\rm ej}$ and the wind magnetization $\sigma$, except as may result from latitudinal variations in the wind power.  Even latitudinal effects are, however, suppressed by the formation of a MWN downstream of the termination shock.  In this region the wind is decelerated, plasma is heated, the magnetic field is compressed, and the pressure settles into a quasi-hydrostatic equilibrium, which depends only on $\sigma$.

In Figure \ref{fig:jets} we show the velocity and density profiles at $t = 35$ seconds from calculations performed for three different values of the ratio $\dot{E}/M_{\rm ej} = 10^{50}, 10^{51}, $and $ 10^{52}$ erg s$^{-1}$ M$_\odot^{-1}$ (cases A, B, and C, respectively).  Significant differences are immediately apparent in the properties of the outflow and the overall dynamics of the MWN-ejecta interaction between the three cases.  In both the low power (A) and `average' (B) cases the MWN is confined within the ejecta.  A well-collimated bipolar relativistic outflow develops, qualitatively similar to that found in the core collapse context as applied to LGRB (e.g.~\citealt{bucciantini08}).  In contrast, in the high power case (C) the MWN has almost completely blown the shell apart. 

Quantifying the geometry of the jet, in order to measure e.g.~the jet opening angle, is nontrivial because its shape is not simply conical.  In case A the jet shape is parabolic, while in case C the jet `flares out' into a diverging flow.  In the high power case C, it is unclear whether the ejecta will provide any confinement at all.  Although Rayleigh-Taylor instability is fundamentally a three-dimensional process, and in principle axisymmetric simulations might fail to reproduce properly its detailed growth and geometrical properties, 2D simulations in the context of Pulsar Wind Nebulae \citep{jun98,bucciantini+04} agree with observations in term of the size and average properties of the unstable mixing layer. This might be due partly to the presence of a strong toroidal field which can suppress the growth in the azimuthal direction. We note, moreover, that the jet propagates through the ejecta approximately an order of magnitude faster than the time required for the shell to fragment.  Thus, even in the case of an energetic wind, a collimated outflow may form initially in the polar region  (albeit with a wide opening angle).  In principle limited confinement could be hence maintained for a short time $\sim 10-30$ s, before shell fragmentation completes and the outflow becomes more isotropic. 

\begin{figure}
\resizebox{\hsize}{!}{\includegraphics{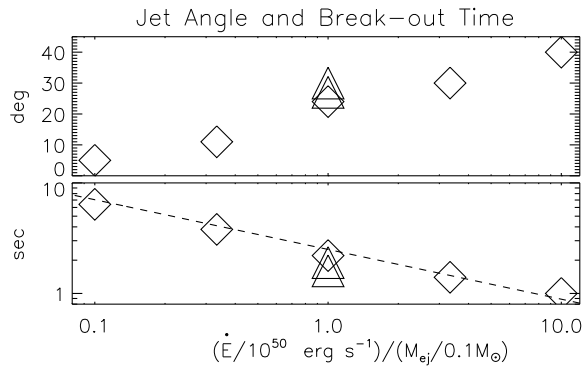}}
\caption{Properties of the relativistic jet.  The upper panel shows the opening angle of the jet $\theta_{\rm j}$ at $t\simeq 35$ s (Fig.~\ref{fig:jets}) as a function of the ratio $\dot{E}/M_{\rm ej}$.  Diamonds represent cases for which the ejecta shell is uniform with angle, while triangles show cases assuming a lower polar densities ($\alpha =3-5$; see eq.~[\ref{eq:rho}]). The lower panel shows the time that the jet `breaks out' from the shell of ejecta (symbols are the same as in the upper panel).  The dashed line is a power-law with exponent $0.5$.}
\label{fig:and}
\end{figure}

Figure ~\ref{fig:and} shows the `break-out' time and characteristic opening angle of the jet as a function of $\dot{E}/M_{\rm ej}$, calculated from several simulations including those shown in Figure \ref{fig:jets}.  The break-out time scales approximatively as $\propto (\dot{E}/M_{\rm ej})^{-1/2}$.  Although, given the self-similar nature of the ejecta, we expect that quantities should depend primarily on the ratio $\dot{E}/M_{\rm ej}$, the precise functional dependence is non-trivial to derive.  Although we find that the basic jet properties are relatively robust to our assumed value for $C$, more substantial changes could in principle result for different values of $\dot{E}/M_{\rm j}$, $r_{\rm e}$ and $r_{\rm in}$.  Nevertheless, we do not expect large variations in the latter quantities, with respect to the fiducial values adopted in this paper. 

Extrapolating Figure \ref{fig:and} to low values $\dot{E}/M_{\rm ej} \lesssim 10^{49}$ erg s$^{-1}$ M$_\odot^{-1}$, we find that the jet requires $\gtrsim 20$ s to break out.  This timescale is comparable to both the delay observed before the onset of the EE in SGRBEEs, and to the time that the  proto-magnetar wind spends at its highest spin-down luminosity \citep{metzger10}.  $\dot{E}/M_{\rm ej} \sim 10^{49}$ erg s$^{-1}$ M$_\odot^{-1}$ thus represents a reasonable threshold power below which the jet is `choked' inside the ejecta, in which case GRB-like high energy emission would fail.  Naively it may appear that a similar criterion should apply also to the case of LGRBs, for which the typical mass of the surrounding envelope $\sim 10$M$_\odot$ would imply a limit on the wind power $\gtrsim 10^{50}$erg s$^{-1}$.  Note, however, several important differences between the dynamics of the MWN in the LGRB and SGRBEE contexts.  In the SGRBEE case, the ejecta expands at a significant fraction of the speed of light, such that at $t \sim 20$ s the shell radius ($\sim 10^{11}$ cm) is significantly greater than the $\sim 10^{10}$ cm radius of the progenitor envelope in the core collapse case.  Since the polar expansion of the MWN is driven by internal pressure, the MWN suffers larger adiabatic losses in the SGRBEE scenario.  The true threshold for a choked jet in the LGRB case thus occurs at lower luminosities $\gtrsim 10^{49}$ erg s$^{-1}$, more consistent with previous estimates in the literature \citep{matzner03}.  

Figures~\ref{fig:jets} and \ref{fig:and} show that the opening angle of the jet $\theta_{\rm j}$ increases with $\dot{E}/M_{\rm ej}$.  One reason for this dependence is that $\theta_{\rm j}$ is proportional to the ratio of the pressure scale height of the MWN in the cylindrical direction $H$ (resulting from magnetic stresses perpendicular to the vertical axis) to the radius of the ejecta $r_{\rm e}$.  Since $H$ is itself proportional to the radius of the MWN \citep{bucciantini07}, which is larger for more energetic jets, it follows that $\theta_{\rm j}$ also increases with $\dot{E}/M_{\rm ej}$.

In addition to the calculations performed for isotropic ejecta (as in Figure \ref{fig:jets}), in Figure \ref{fig:and} we also show results for cases with a lower polar density ($\alpha=3-5$ in eq.~[\ref{eq:rho}]).  We find that the jet break-out times and opening angles show a similar dependence to the isotropic case, provided that our results are parameterized in terms of an effective ejecta mass, defined as that of a spherically symmetric shell with a density equal to the polar value.  

We now briefly discuss the acceleration and variability of the jet.  In all cases we find that the jet accelerates approximately linearly with radius (i.e. bulk Lorentz factor $\gamma \propto r$), suggesting efficient acceleration, as in traditional GRB fireball models (e.g.~\citealt{Goodman86,Paczynski86}), although acceleration is slightly faster in the lower power case (A).  This is consistent with previous studies of relativistic jets, for which it is found that acceleration is more effective for more narrowly collimated outflows \citep{komissarov10}.  We also find that the Lorentz factor of the jet has a `top-hat' angular profile, with a constant-$\gamma$ core and a sharp velocity gradient at the boundary to $\gamma \gtrsim 1$.   

Although the properties of the injected magnetar wind are held constant in time, the properties of the jet as it leaves the ejecta are nevertheless variable.  In particular, the Lorentz factor experiences order-unity fluctuations in the less energetic (more collimated) cases A and B on a lengthscale of $\sim 2\times 10^{11}$ cm, corresponding to a typical timescale $\sim 5-6 $ sec.  In the high power case C the jet properties show a smoother time evolution.  One reason for this difference is that variability results from interaction in the magnetar nebula (in particular with the walls of the ejecta channel; \citealt{Morsony+10}): higher power jets are less variable because the ejecta channel is larger and provides a less-collimating environment.  Indeed, we find that the observed timescales are similar to the sound/Alfv\'enic time across the jet.  Although we cannot rule out shorter timescale variability, as it would not be resolved, significant changes in the fluid geometry on small scales seem unlikely given the relatively large size of the MWN system.  As we discuss in Section \ref{sec:discussion}, because narrow jets are predicted to be more variable than wider ones, the measured large amplitude variability of the EE provides an independent diagnostic of the jet opening angle. 

Because the proto-magnetar model for GRBs predicts a positive correlation between $\dot{E}$ and the wind magnetization $\sigma$ \citep{Thompson+04,metzger10}, we also repeat the case B calculation for a higher value $\sigma = 0.2$.  We find that the jet break-out time is similar between the low and high magnetization cases, but the opening angle is a few degrees smaller in the high-$\sigma$ case.  We want to stress here that we assume efficint conversion of magnetic to kinetic energy either in the wind or at the termination shock such that the system behaves as if $\sigma < 1$ even for outflows that are Poyting flux dominated at the Ligth Cylinder [see the discussion in \citet{bucciantini07} for the case of a proto-neutron star wind in the context of LGRBs]. The problem of magnetic dissipation in winds has been investigated in the case of pulsars \citep{Kirk&Skjaerasen03}, while for for dissipation at the termination shock  the reader is referred to the recent results by \citet{Sironi&Spitkovsky11}. Was also repeated our case B calculation for a shallower density profile $C=2$ (eq.~[\ref{eq:rho}]).  Again we found no significant difference with respect to the reference case $C=4$, which suggests that our results do not depend sensitively on the precise structure of the ejecta.

\section{Implications for SGRBEEs}
\label{sec:discussion}

\begin{figure}
\resizebox{\hsize}{!}{\includegraphics{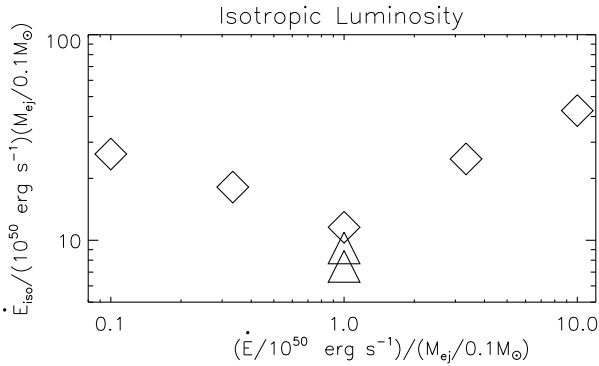}}
\caption{Isotropic jet luminosity $\dot{E}_{\rm iso} \equiv \dot{E}f_{\rm b}^{-1}$, where $f_{\rm b} = \Omega/4\pi \approx \theta_{\rm j}^{2}/2$ is the jet beaming fraction and $\Omega$ is the opening solid angle of the jet, after it has relaxed on a timescale $t\simeq 35$ s (Fig.~\ref{fig:jets}).  Diamonds show calculations performed assuming uniform ejecta, while triangles show cases with a lower polar density. }
\label{fig:liso}
\end{figure}

After a clean opening is created through the ejecta, the power of the jet reflects, in a time- and angle-averaged sense, the value of $\dot{E}(t)$ injected by the proto-magnetar wind at much smaller radii (e.g.~\citealt{bucciantini09}; \citealt{Morsony+10}).  Most emission models for GRBs, such as internal shocks or magnetic reconnection, are the result of internal dissipation in the jet.  In this case the observed (photon) luminosity is typically proportional to the beaming-corrected, or {\it isotropic}, jet luminosity $\dot{E}_{\rm iso} \equiv \dot{E}f_{\rm b}^{-1}$, where $f_{\rm b} \simeq \theta_{\rm j}^{2}/2$ is the beaming fraction.  

Figure \ref{fig:liso} shows $\dot{E}_{\rm iso}$ as a function of $M_{\rm ej}/\dot{E}$ for the same calculations shown in Figure \ref{fig:and}.  The inverse correlation between the wind power and opening angle (Fig.~\ref{fig:and}) implies that, at fixed ejecta mass, $\dot{E}_{\rm iso}$ is constant to within a factor $\sim 3$ across two orders of magnitude in $\dot{E}$.  Thus, as long as the shell effectively collimates the outflow, the EE luminosity in the proto-magnetar model depends primarily on $M_{\rm ej}$.   If, by contrast, $\dot{E}/M_{\rm ej} > 10^{52}$ erg s$^{-1}$ M$_\odot^{-1}$ then the shell may be entirely disrupted (in which case the isotropic luminosity is instead directly $\propto \dot{E}$), while if $\dot{E}/M_{\rm ej} < 10^{49}$ erg s$^{-1}$ M$_\odot^{-1}$ the jet is probably choked and no emission is expected on timescales of relevance.  

The geometry of the magnetar jet also has consequences for the ubiquity of EE associated with short GRBs.  As noted by \citet{metzger08}, without confinement the magnetar outflow (responsible for the EE) is mostly equatorial, while the accretion-powered jet (responsible for the initial short GRB) is probably polar.  An important question is thus whether a typical observer will see both components.  Our results show that, except perhaps in the most energetic cases, the magnetar wind is diverted into a polar outflow.  Unfortunately, the opening angles of SGRBs are poorly constrained\footnote{In such a `two jet' scenario it is also unclear {\it which} jet to associate a putative opening angle measurement with (e.g.~\citealt{Granot05}).} observationally, with measured values ranging from a few to $>25$ degrees \citep{burrows06,grupe06}. Recent numerical results suggest values in the range of $\sim 20$ degree \citep{Rezzolla+11}. It is thus possible that events could exist for which the extended emission is not observable because it is more collimated than the initial SGRB, in which case the event would be classified as a `normal' short burst (see \citealt{Barkov&Pozanenko11} for a similar idea).  Such events cannot be too common because the fraction of short GRBs with observed EE is already rather large \citep{Norris&Bonnell06}.  This implies that the magnetar wind cannot be too collimating, which suggests that the average shell mass is low (we provide additional evidence for a wide-angle magnetar jet below).  

Alternatively, events may exist for which only the EE is observable, because the initial short burst is more narrowly collimated.  These events would probably be classified as regular long duration GRBs or X-ray Flashes, but would not be accompanied by a bright associated supernova.  It is difficult to place definitive constaints on the rate of such events, although we note that at least one X-ray Flash with an EE-like light curve was not in fact accompanied by a bright supernova (XRF 040701; \citealt{Soderberg+05}). 

\begin{figure}
\resizebox{\hsize}{!}{\includegraphics{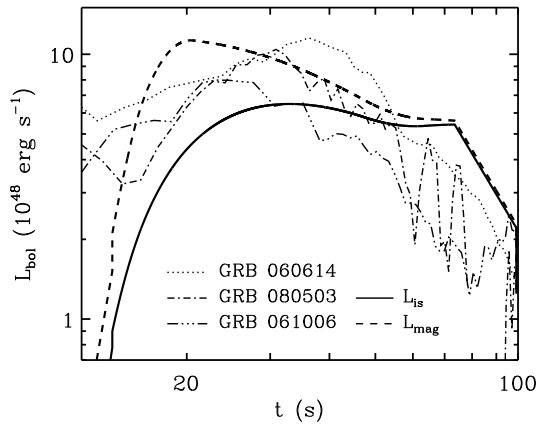}}
\caption{Average bolometric luminosity of GRB emission from the proto-magnetar jet as a function of time after formation, calculated using the models described in \citet{metzger10}.  Emission predicted by the internal shock and magnetic dissipation models are shown with solid and dashed lines, respectively.  The calculation assumes that the magnetar has an aligned dipole field of strength $B_{\rm dip} = 2\times 10^{15}$ G and an initial spin period $P_{0} = 1.5$ ms.  We adopt a value for the electron radiative efficiency $\epsilon_{e} = 0.2$ and a beaming fraction $f_b = 0.3$ (see text).  For comparison we also plot the $15-350$ keV {\it Swift} BAT extended emission light curves for GRBs 060614 ({\it dotted}), 080503 ({\it dot-dashed}), and 061005 ({\it triple-dot-dashed}) \citep{Butler&Kocevski07}.   
}
\label{fig:lcs}
\end{figure}

We now attempt to constrain the properties of the ejecta using the measured luminosity of the EE.  The sample of SGRBEEs with known redshifts and measured EE fluences is unfortunately small and incomplete.  The sample may furthermore be biased against less luminous events, in which case the lower limits are not constraining.  Nevertheless, when measured, the isotropic luminosity of the EE is typically in the  range $L_{\rm EE} \sim 2\times 10^{48} - 2\times 10^{49}$ erg s$^{-1}$ (Figure \ref{fig:lcs} shows some examples).  Since we found that during the early jet-formation phase the isotropic luminosity is  (Fig.~\ref{fig:liso})
\begin{equation}
\dot{E}_{\rm iso,j} \sim 1-3\times 10^{51}  (M_{\rm ej}/0.1{\rm M}_\odot)  {\rm erg~ s}^{-1},
\end{equation}
we can relate the observed EE luminosity $L_{\rm EE}$ to the ejecta mass:
\be
M_{\rm ej} \sim 0.01-0.03\left(\frac{L_{\rm EE}}{10^{49}{\,\rm erg\,s^{-1}}}\right)\left(\frac{\eta_{\rm p}}{0.1}\right)^{-1}\left(\frac{\eta_{\rm rad}}{0.3}\right)^{-1}M_{\sun},
\label{eq:mej}
\ee
where $\eta_{\rm rad} \equiv L_{\rm EE}/\dot{E}_{\rm iso,EE}$ is the radiative efficiency of the jet, and $\eta_{\rm p} \equiv \dot{E}_{\rm iso,EE}/\dot{E}_{\rm iso, j} \sim 0.1-0.3$ is the ratio between the isotropic power of the magnetar wind during the EE phase $\dot{E}_{\rm iso,EE}$ at late times ($t \sim 10-100$ s) and that at early times $\dot{E}_{\rm iso,j}$ ($t \lesssim 10$ s), when the opening angle of the jet is determined.  Detailed evolutionary models of proto-magnetar spin-down \citep{metzger10} show that $\eta_{\rm p}$ is typically $\sim 0.1-0.3$.

Equation (\ref{eq:mej}) shows that for typical values of $\eta_{\rm rad}$, $\eta_{\rm EE}$, and the measured EE luminosity, the inferred ejecta masses are in the range $M_{\rm ej} \sim 10^{-3}-10^{-1}M_{\sun}$, consistent with the range of predicted ejecta masses in both the NS-NS merger and AIC scenarios (Sec.~\ref{sec:intro}).  This represents an important consistency check on the proto-magnetar model.
 
Adopting a typical value of the ejecta mass $M_{\rm ej} = 10^{-2}M_{\sun}$, our results require that the early-phase power of the proto-magnetar wind must lie in the range $\dot{E}\sim 10^{48}-10^{50}$ ergs s$^{-1}$ so as to both maintain confinement and jet collimation, yet not produce a choked jet.  This spin-down luminosity corresponds to magnetars with either a relatively slow initial rotation rate (spin period $P_{0} \gtrsim 3$ ms) or a relatively weak dipole magnetic field strength $B_{\rm dip} \lesssim 2\times 10^{15}$ G (Metzger et al.~2010), as compared to those required to produce a bright Classical GRB.  However, rotation periods much larger than the break-up rate ($P_{0} \sim 1-2$ ms) seem unlikely given the substantial angular momentum that is required in both NS-NS merger and AIC scenarios to form an initial disk.  Thus, we are led to consider somewhat lower field magnetars.

In Figure \ref{fig:lcs} we show models for the average bolometric luminosity of GRB emission from the proto-magnetar jet, calculated using the models described in \citet{metzger10}.  We show the predicted emission in both internal shock and magnetic dissipation models, assuming the magnetar has an aligned dipole field of strength $B_{\rm dip} = 2\times 10^{15}$ G and an initial spin period $P_{0} = 1.5$ ms.  For comparison, we also plot the extended emission light curves of several {\it Swift}-detected SGRBEEs.  In this model the jet power at $t \sim 1-10$ s is $\dot{E} \sim 10^{49}-10^{50}$ erg s$^{-1}$.  Applying these values to Figure \ref{fig:and}, we adopt a beaming fraction $f_{\rm b} \sim 0.3$ corresponding to a relatively large opening angle $\theta_{\rm j} \sim 45^{\circ}$.  

Figure \ref{fig:lcs} demonstrates that the emission predicted by the proto-magnetar model qualitatively agrees with the observed onset, duration, and luminosity of SGRBEEs.  Also note the presence of a phase of steep decay in some of the SGRBEE light curves at late times.  A relatively abrupt shut off in prompt emission is indeed expected to occur in the magnetar model when the (initially opaque) proto-neutron star becomes transparent to neutrino emission on a timescale $\sim 30-100$ s (see \citealt{metzger10} for a discussion).  The observed spectral properties of SGRBEEs can also be compared to the predictions of our emission models; in GRB 060614, for instance, the spectrum peaked at an energy $E_{\rm peak} \sim 50$ keV, which decreased by a factor $\sim 2$ throughout the EE (e.g.~\citealt{Zhang+07}).  For internal shock emission, the value of $E_{\rm peak}$ and its time evolution are unfortunately very sensitive functions of the assumed microphysical parameters, which makes a definitive comparison to our model challenging.  By contrast, the magnetic dissipation model indeed predicts a lower value $E_{\rm peak} \lesssim 100$ keV for jets with smaller values of $\dot{E}_{\rm iso}$ and $\sigma$ (see \citealt{metzger10}; their eq.~[11]) as compared to those associated with standard long duration GRBs.

Although Figure \ref{fig:lcs} shows a reasonably good fit to the EE light curves, similar agreement would not obtain for lower-$B_{\rm dip}$ magnetars with less powerful winds.  Less powerful outflows are `dirtier' and require at least several tens of seconds before becoming optically-thin at the emission radius (see Fig.~8 of \citealt{metzger10}), inconsistent with the much earlier observed onset of the EE.  Because the magnetar wind cannot be too weak, self-consistency with the results of Figure \ref{fig:and} demands that the jet opening angle cannot be too small. We want to stress here that the problem of relating the outflow kinetic luminosity with the observed $\gamma$-ray emission, depends on assumptions about acceleration efficiency and radiation mechanism that are still poorely undestood, even in the canonical collpasar model of LGRBs. The present resuls should be considered indicative that the model can in principle reproduce the data at least in their broad temporal evolution. Indeed, the magnetar wind could also be much more powerful powerful ($\dot{E} \gg 10^{50}$ ergs s$^{-1}$), in which case the shell may be completely blown apart (case C in Fig.~\ref{fig:jets}).  As discussed above, in this case the rate of events for which the EE would be observable without an accompanying short GRB would be quite high.  Future all sky X-ray missions (e.g.~JANUS; \citealt{Fox+10}) could detect SGRBEEs by triggering on their EE alone, and hence may capable of addressing this question.  

An independent way to break the degeneracy between wide and narrow-angle jets is with variability.  As discussed in Sect.~\ref{sec:results}, the amplitude of fluctuations in the jet properties scale inversely with the jet opening angle.  Since EE are generally less variable than standard long duration GRBs, and the latter are estimated to have opening angles in the range of a few-10 degrees (e.g.~\citealt{Frail+01}), this also hints that larger opening angles are favoured for SGRBEEs.  

\section{Conclusions}
\label{sec:conclusions}

The long durations and other similarities of the extended emission following some short GRBs is difficult to understand within the standard NS merger scenario.  Building on the work of \citet{metzger08}, in this paper we present numerical simulations of the interaction between the energetic wind from a remnant proto-magnetar and the ejecta from the merger or AIC event.  We analyze the confining properties of the ejecta shell and calculate the dependence of the jet properties on the properties of the proto-magnetar wind.

Our results show that, for the self-similar ejecta structure that we have assumed, the evolution of the system is primarily a function of the ratio between the power of the magnetar wind and the mass of the ejecta.  Latitudinal density variations can be reparametrized in term of an effective mass.  We also demonstrate that well-collimated jets form only over a limited range in energy: a sufficiently powerful wind completely disrupts the confining shell, while a lower wind instead produces a choked jet.  Interestingly, we find an anti-correlation between the wind power and opening angle (in the range of $\dot{E}/M_{\rm ej}$ over which a collimated jet forms), which implies that the observed (isotropic) luminosity is almost independent of $\dot{E}/M_{\rm ej}$ (Fig.~\ref{fig:liso}).  Adopting a fiducial shell mass $M_{\rm ej} \sim 10^{-2}M_{\sun}$, the corresponding luminosity $\sim 10^{49}$ ergs s$^{-1}$ is broadly consistent with observed EE luminosities (Fig.~\ref{fig:lcs}).  

Considering the wind properties necessary to produce bright non-thermal emission also constrains the types of magnetars capable of powering the EE.  Only magnetars with relatively strong surface dipole fields $B_{\rm dip} \gtrsim 10^{15}$ G produce winds that are sufficiently `clean', such that they become optically thin at the radii where the jet energy is dissipated sufficiently early to explain the $\sim 10$ s observed onset of the EE.  Consistent with this notion, we find that a millisecond magnetar with $\gtrsim 10^{15}$ G indeed produces emission qualitatively consistent with observed EE light curves (Fig.~\ref{fig:lcs}).

Our investigation has primarily focused on a fiducial profile for the ejecta.  As discussed in the Introduction, although the ejecta mass may vary by a factor $\gtrsim 100$ depending on e.g.~the scenario responsible for producing the magnetar (Fig.~\ref{fig:cartoon}), the characteristic size and velocity of the ejecta are not expected to vary by more than a factor of a few.  For these reasons our results may be relatively robust.  We caution, however, that a full parameter study is necessary to properly investigate questions such as to what degree the opening angle and break-out time depend on additional parameters, such as the relative size of the inner and outer radius of the ejecta.
 
One limitation of our calculations is that the properties of the jet depend on the assumed magnetization of the wind $\sigma = 0.1$, which was chosen because (1) previous results in the context of LGRBs \citep{bucciantini08} agree with more detailed calculations where the magnetization in the wind evolves according to the value near the light cylinder predicted from evolutionary models including PNS cooling \citep{bucciantini09}; and (2) because a variety of processes at work in the wind, the termination shock, and the nebula, may dissipate the toroidal flux, as is inferred in pulsar wind nebulae.  Recently \citet{mizuno11} have shown that current driven instability can efficiently reduce the amount of toroidal magnetic field. They conclude that the magnetic field in the nebula can behave as if the effective $\sigma$ is much lower (less by a factor 10) than the true value.  If this also holds in the more complex scenario of a MWN, then it is possible that current-driven instabilities may regulate the magnetization, such that the lower effective value of $\sigma$ as we have chosen is appropriate.

Up to this point we have not discussed how to distinguish between proto-magnetars formed from AIC and via NS-NS mergers, since the physical scenario is similar (see Fig.~\ref{fig:cartoon}).  Although in neither case do we expect the event to be associated with bright supernova, the ejecta is nevertheless composed of radioactive isotopes.  As this material decays to stability, it reheats the ejecta, powering transient optical emission lasting $\sim 1$ day, with a peak brightness $\nu L_{\nu} \sim 10^{41-42}$ ergs s$^{-1}$, approximately a thousand times brighter than a nova (\citealt{Metzger+09b}; \citealt{Metzger+10b}).  Although such `kilonova' have not yet been definitively detected (although see \citealt{perley09}), current searches are underway (\citealt{Kasliwal+10}).  One way to distinguish between AIC and NS-NS mergers is that, in AIC the ejecta is predicted to be rich in Fe-group elements (e.g.~$^{56}$Ni; \citealt{Metzger+09b}; \citealt{darbha10}), while in NS-NS mergers the ejecta is composed of more exotic, heavy $r$-process nuclei (\citealt{Metzger+10b}).  We note that if the proto-magnetar injects additional energy behind the expanding ejecta, and if this energy is thermalized, this could enhance the luminosity of the predicted kilonova emission, as in analogous models for very luminous core collapse supernova (\citealt{Kasen&Bildsten10}; \citealt{Woosley10}; \citealt{Piro&Ott11}).  We plan to explore this possibility further in future work. Evidence of a large opening angle for magnetar outflows in SGRBEEs suggests the possibility of a strong interaction between the flow and (1) the companion in the AIC scenario, (2) a possible tertiary component in the NS-NS merger
scenario \citep{Thompson10}, or (3) the clumps of matter ejected in unbound orbits in both scenarios. Such interactions could lead to observational signatures, which might help discriminate among the various scenarios ({\citealt{marietta+00}; \citealt{macfadyen+05}).

Another characteristic distinguishing NS-NS mergers and AIC is the strength and form of their gravitational wave signal \citep{abdikalamov10}.  The coincident detection (or constraining upper limits) on gravitational wave emission from SGRBEEs with future detectors such as Advanced LIGO/Virgo should conclusively settle this issue.  The detection of a giant SGR-like flare from the location of a previous SGRBEE \citep{Giannios10} would be `smoking gun' proof that magnetar birth is at the origin of these events.

\section*{Acknowledgments}
We thank Daniel Perley for his assistance with the {\it Swift} data.  NB is supported by a NORDITA fellowship grant.  BDM is supported by
NASA through Einstein Postdoctoral Fellowship grant number PF9-00065
awarded by the Chandra X-ray Center, which is operated by the
Smithsonian Astrophysical Observatory for NASA under contract
NAS8-03060. TAT is supported in part by an Alfred P. Sloan Foundation
Fellowship.

\vspace{-0.cm}


\bibliography{ms2.bib}{}

\begin{thebibliography}{111}
\expandafter\ifx\csname natexlab\endcsname\relax\def\natexlab#1{#1}\fi

\bibitem[{{Abdikamalov} {et~al}\mbox{.}(2010){Abdikamalov}, {Ott}, {Rezzolla},
  {Dessart}, {Dimmelmeier}, {Marek}, \& {Janka}}]{abdikalamov10}
{Abdikamalov} E.~B., {Ott} C.~D., {Rezzolla} L., {Dessart} L., {Dimmelmeier}
  H., {Marek} A., {Janka} H., 2010, Phy.Rev.D, 81, 044012

\bibitem[{{Akiyama} {et~al}\mbox{.}(2003){Akiyama}, {Wheeler}, {Meier}, \&
  {Lichtenstadt}}]{Akiyama+03}
{Akiyama} S., {Wheeler} J.~C., {Meier} D.~L., {Lichtenstadt} I., 2003, \apj,
  584, 954

\bibitem[{{Baiotti}, {Giacomazzo} \& {Rezzolla}(2008){Baiotti}, {Giacomazzo},
  \& {Rezzolla}}]{baiotti+08}
{Baiotti} L., {Giacomazzo} B., {Rezzolla} L., 2008, \prd, 78, 084033

\bibitem[{{Barkov} \& {Pozanenko}(2011)}]{Barkov&Pozanenko11}
{Barkov} M.~V., {Pozanenko} A.~S., 2011, ArXiv e-prints

\bibitem[{{Barthelmy} {et~al}\mbox{.}(2005){Barthelmy} {et~al.}}]{Barthelmy+05}
{Barthelmy} S.~D., {et~al.}, 2005, \nat, 438, 994

\bibitem[{{Baumgarte}, {Shapiro} \& {Shibata}(2000){Baumgarte}, {Shapiro}, \&
  {Shibata}}]{Baumgarte+00}
{Baumgarte} T.~W., {Shapiro} S.~L., {Shibata} M., 2000, \apjl, 528, L29

\bibitem[{{Begelman}(1998)}]{begelman98}
{Begelman} M.~C., 1998, \apj, 493, 291

\bibitem[{{Begelman} \& {Li}(1992)}]{Begelman&Li92}
{Begelman} M.~C., {Li} Z., 1992, \apj, 397, 187

\bibitem[{{Berger}(2009)}]{berger09}
{Berger} E., 2009, \apj, 690, 231

\bibitem[{{Berger} {et~al}\mbox{.}(2005){Berger} {et~al.}}]{Berger+05}
{Berger} E., {et~al.}, 2005, \nat, 438, 988

\bibitem[{{Bloom}, {Butler} \& {Perley}(2008){Bloom}, {Butler}, \&
  {Perley}}]{Bloom+08}
{Bloom} J.~S., {Butler} N.~R., {Perley} D.~A., 2008, in American Institute of
  Physics Conference Series, Vol. 1000, American Institute of Physics
  Conference Series, {M.~Galassi, D.~Palmer, \& E.~Fenimore}, ed., pp. 11--15

\bibitem[{{Bloom} {et~al}\mbox{.}(2006){Bloom}, {Prochaska}, {Pooley}, {Blake},
  {Foley}, {Jha}, {Ramirez-Ruiz}, {Granot}, {Filippenko}, {Sigurdsson},
  {Barth}, {Chen}, {Cooper}, {Falco}, {Gal}, {Gerke}, {Gladders}, {Greene},
  {Hennanwi}, {Ho}, {Hurley}, {Koester}, {Li}, {Lubin}, {Newman}, {Perley},
  {Squires}, \& {Wood-Vasey}}]{bloom06}
{Bloom} J.~S. {et~al.}, 2006, \apj, 638, 354

\bibitem[{{Bucciantini} {et~al}\mbox{.}(2004){Bucciantini}, {Amato},
  {Bandiera}, {Blondin}, \& {Del Zanna}}]{bucciantini+04}
{Bucciantini} N., {Amato} E., {Bandiera} R., {Blondin} J.~M., {Del Zanna} L.,
  2004, \aap, 423, 253

\bibitem[{{Bucciantini} {et~al}\mbox{.}(2007){Bucciantini}, {Quataert},
  {Arons}, {Metzger}, \& {Thompson}}]{bucciantini07}
{Bucciantini} N., {Quataert} E., {Arons} J., {Metzger} B.~D., {Thompson} T.~A.,
  2007, \mnras, 380, 1541

\bibitem[{{Bucciantini} {et~al}\mbox{.}(2008){Bucciantini}, {Quataert},
  {Arons}, {Metzger}, \& {Thompson}}]{bucciantini08}
---, 2008, \mnras, 383, L25

\bibitem[{{Bucciantini} {et~al}\mbox{.}(2009){Bucciantini}, {Quataert},
  {Metzger}, {Thompson}, {Arons}, \& {Del Zanna}}]{bucciantini09}
{Bucciantini} N., {Quataert} E., {Metzger} B.~D., {Thompson} T.~A., {Arons} J.,
  {Del Zanna} L., 2009, \mnras, 396, 2038

\bibitem[{{Bucciantini} {et~al}\mbox{.}(2006){Bucciantini}, {Thompson},
  {Arons}, {Quataert}, \& {Del Zanna}}]{Bucciantini+06}
{Bucciantini} N., {Thompson} T.~A., {Arons} J., {Quataert} E., {Del Zanna} L.,
  2006, \mnras, 368, 1717

\bibitem[{{Burrows} {et~al}\mbox{.}(2006){Burrows}, {Grupe}, {Capalbi},
  {Panaitescu}, {Patel}, {Kouveliotou}, {Zhang}, {M{\'e}sz{\'a}ros},
  {Chincarini}, {Gehrels}, \& {Wijers}}]{burrows06}
{Burrows} D.~N. {et~al.}, 2006, \apj, 653, 468

\bibitem[{{Butler} \& {Kocevski}(2007)}]{Butler&Kocevski07}
{Butler} N.~R., {Kocevski} D., 2007, \apj, 663, 407

\bibitem[{{Camus} {et~al}\mbox{.}(2009){Camus}, {Komissarov}, {Bucciantini}, \&
  {Hughes}}]{camus09}
{Camus} N.~F., {Komissarov} S.~S., {Bucciantini} N., {Hughes} P.~A., 2009,
  \mnras, 400, 1241

\bibitem[{{Chawla} {et~al}\mbox{.}(2010){Chawla}, {Anderson}, {Besselman},
  {Lehner}, {Liebling}, {Motl}, \& {Neilsen}}]{Chawla+10}
{Chawla} S., {Anderson} M., {Besselman} M., {Lehner} L., {Liebling} S.~L.,
  {Motl} P.~M., {Neilsen} D., 2010, Physical Review Letters, 105, 111101

\bibitem[{{Chornock} {et~al}\mbox{.}(2010){Chornock}, {Berger}, {Levesque},
  {Soderberg}, {Foley}, {Fox}, {Frebel}, {Simon}, {Bochanski}, {Challis},
  {Kirshner}, {Podsiadlowski}, {Roth}, {Rutledge}, {Schmidt}, {Sheppard}, \&
  {Simcoe}}]{chornock10}
{Chornock} R. {et~al.}, 2010, ArXiv e-prints

\bibitem[{{Darbha} {et~al}\mbox{.}(2010){Darbha}, {Metzger}, {Quataert},
  {Kasen}, {Nugent}, \& {Thomas}}]{darbha10}
{Darbha} S., {Metzger} B.~D., {Quataert} E., {Kasen} D., {Nugent} P., {Thomas}
  R., 2010, \mnras, 409, 846

\bibitem[{{Del Zanna}, {Bucciantini} \& {Londrillo}(2003){Del Zanna},
  {Bucciantini}, \& {Londrillo}}]{delzanna03}
{Del Zanna} L., {Bucciantini} N., {Londrillo} P., 2003, \aap, 400, 397

\bibitem[{{Del Zanna} {et~al}\mbox{.}(2007){Del Zanna}, {Zanotti},
  {Bucciantini}, \& {Londrillo}}]{delzanna07}
{Del Zanna} L., {Zanotti} O., {Bucciantini} N., {Londrillo} P., 2007, \aap,
  473, 11

\bibitem[{{Della Valle} {et~al}\mbox{.}(2006){Della Valle}, {Malesani},
  {Bloom}, {Benetti}, {Chincarini}, {D'Avanzo}, {Foley}, {Covino}, {Melandri},
  {Piranomonte}, {Tagliaferri}, {Stella}, {Gilmozzi}, {Antonelli}, {Campana},
  {Chen}, {Filliatre}, {Fiore}, {Fugazza}, {Gehrels}, {Hurley}, {Mirabel},
  {Pellizza}, {Piro}, \& {Prochaska}}]{dellavalle06}
{Della Valle} M. {et~al.}, 2006, \apjl, 642, L103

\bibitem[{{Demorest} {et~al}\mbox{.}(2010){Demorest}, {Pennucci}, {Ransom},
  {Roberts}, \& {Hessels}}]{Demorest+10}
{Demorest} P.~B., {Pennucci} T., {Ransom} S.~M., {Roberts} M.~S.~E., {Hessels}
  J.~W.~T., 2010, \nat, 467, 1081

\bibitem[{{Dermer} \& {Atoyan}(2006)}]{dermer06}
{Dermer} C.~D., {Atoyan} A., 2006, \apjl, 643, L13

\bibitem[{{Dessart} {et~al}\mbox{.}(2006){Dessart}, {Burrows}, {Ott}, {Livne},
  {Yoon}, \& {Langer}}]{Dessart+06}
{Dessart} L., {Burrows} A., {Ott} C.~D., {Livne} E., {Yoon} S., {Langer} N.,
  2006, \apj, 644, 1063

\bibitem[{{Dessart} {et~al}\mbox{.}(2009){Dessart}, {Ott}, {Burrows},
  {Rosswog}, \& {Livne}}]{Dessart+09}
{Dessart} L., {Ott} C.~D., {Burrows} A., {Rosswog} S., {Livne} E., 2009, \apj,
  690, 1681

\bibitem[{{Duncan} \& {Thompson}(1992)}]{Duncan&Thompson92}
{Duncan} R.~C., {Thompson} C., 1992, \apjl, 392, L9

\bibitem[{{Faber} {et~al}\mbox{.}(2006){Faber}, {Baumgarte}, {Shapiro}, \&
  {Taniguchi}}]{Faber+06}
{Faber} J.~A., {Baumgarte} T.~W., {Shapiro} S.~L., {Taniguchi} K., 2006, \apjl,
  641, L93

\bibitem[{{Fong}, {Berger} \& {Fox}(2010){Fong}, {Berger}, \& {Fox}}]{fong10}
{Fong} W., {Berger} E., {Fox} D.~B., 2010, \apj, 708, 9

\bibitem[{{Fox} {et~al}\mbox{.}(2005){Fox}, {Frail}, {Price}, {Kulkarni},
  {Berger}, {Piran}, {Soderberg}, {Cenko}, {Cameron}, {Gal-Yam}, {Kasliwal},
  {Moon}, {Harrison}, {Nakar}, {Schmidt}, {Penprase}, {Chevalier}, {Kumar},
  {Roth}, {Watson}, {Lee}, {Shectman}, {Phillips}, {Roth}, {McCarthy}, {Rauch},
  {Cowie}, {Peterson}, {Rich}, {Kawai}, {Aoki}, {Kosugi}, {Totani}, {Park},
  {MacFadyen}, \& {Hurley}}]{Fox05}
{Fox} D.~B. {et~al.}, 2005, \nat, 437, 845

\bibitem[{{Fox} \& {JANUS Team}(2010)}]{Fox+10}
{Fox} D.~B., {JANUS Team}, 2010, in Bulletin of the American Astronomical
  Society, Vol.~42, American Astronomical Society Meeting Abstracts \#215, pp.
  481.15--+

\bibitem[{{Frail} {et~al}\mbox{.}(2001){Frail}, {Kulkarni}, {Sari},
  {Djorgovski}, {Bloom}, {Galama}, {Reichart}, {Berger}, {Harrison}, {Price},
  {Yost}, {Diercks}, {Goodrich}, \& {Chaffee}}]{Frail+01}
{Frail} D.~A. {et~al.}, 2001, \apjl, 562, L55

\bibitem[{{Fruchter} {et~al}\mbox{.}(2006){Fruchter} {et~al.}}]{Fruchter+06}
{Fruchter} A.~S., {et~al.}, 2006, \nat, 441, 463

\bibitem[{{Fynbo} {et~al}\mbox{.}(2006){Fynbo}, {Watson}, {Th{\"o}ne},
  {Sollerman}, {Bloom}, {Davis}, {Hjorth}, {Jakobsson}, {J{\o}rgensen},
  {Graham}, {Fruchter}, {Bersier}, {Kewley}, {Cassan}, {Castro Cer{\'o}n},
  {Foley}, {Gorosabel}, {Hinse}, {Horne}, {Jensen}, {Klose}, {Kocevski},
  {Marquette}, {Perley}, {Ramirez-Ruiz}, {Stritzinger}, {Vreeswijk}, {Wijers},
  {Woller}, {Xu}, \& {Zub}}]{fynbo06}
{Fynbo} J.~P.~U. {et~al.}, 2006, \nat, 444, 1047

\bibitem[{{Gal-Yam} {et~al}\mbox{.}(2006){Gal-Yam}, {Fox}, {Price}, {Ofek},
  {Davis}, {Leonard}, {Soderberg}, {Schmidt}, {Lewis}, {Peterson}, {Kulkarni},
  {Berger}, {Cenko}, {Sari}, {Sharon}, {Frail}, {Moon}, {Brown}, {Cucchiara},
  {Harrison}, {Piran}, {Persson}, {McCarthy}, {Penprase}, {Chevalier}, \&
  {MacFadyen}}]{Gal-Yam+06}
{Gal-Yam} A. {et~al.}, 2006, \nat, 444, 1053

\bibitem[{{Galama} {et~al}\mbox{.}(1998){Galama}, {Vreeswijk}, {van Paradijs},
  {Kouveliotou}, {Augusteijn}, {B{\"o}hnhardt}, {Brewer}, {Doublier},
  {Gonzalez}, {Leibundgut}, {Lidman}, {Hainaut}, {Patat}, {Heise}, {in't Zand},
  {Hurley}, {Groot}, {Strom}, {Mazzali}, {Iwamoto}, {Nomoto}, {Umeda},
  {Nakamura}, {Young}, {Suzuki}, {Shigeyama}, {Koshut}, {Kippen}, {Robinson},
  {de Wildt}, {Wijers}, {Tanvir}, {Greiner}, {Pian}, {Palazzi}, {Frontera},
  {Masetti}, {Nicastro}, {Feroci}, {Costa}, {Piro}, {Peterson}, {Tinney},
  {Boyle}, {Cannon}, {Stathakis}, {Sadler}, {Begam}, \& {Ianna}}]{galama98}
{Galama} T.~J. {et~al.}, 1998, \nat, 395, 670

\bibitem[{{Gehrels} {et~al}\mbox{.}(2006){Gehrels}, {Norris}, {Barthelmy},
  {Granot}, {Kaneko}, {Kouveliotou}, {Markwardt}, {M{\'e}sz{\'a}ros}, {Nakar},
  {Nousek}, {O'Brien}, {Page}, {Palmer}, {Parsons}, {Roming}, {Sakamoto},
  {Sarazin}, {Schady}, {Stamatikos}, \& {Woosley}}]{Gehrels+06}
{Gehrels} N. {et~al.}, 2006, \nat, 444, 1044

\bibitem[{{Giannios}(2010)}]{Giannios10}
{Giannios} D., 2010, \mnras, 403, L51

\bibitem[{{Goodman}(1986)}]{Goodman86}
{Goodman} J., 1986, \apjl, 308, L47

\bibitem[{{Granot}(2005)}]{Granot05}
{Granot} J., 2005, \apj, 631, 1022

\bibitem[{{Grupe} {et~al}\mbox{.}(2006){Grupe}, {Burrows}, {Patel},
  {Kouveliotou}, {Zhang}, {M{\'e}sz{\'a}ros}, {Wijers}, \& {Gehrels}}]{grupe06}
{Grupe} D., {Burrows} D.~N., {Patel} S.~K., {Kouveliotou} C., {Zhang} B.,
  {M{\'e}sz{\'a}ros} P., {Wijers} R.~A.~M., {Gehrels} N., 2006, \apj, 653, 462

\bibitem[{{Hjorth} {et~al}\mbox{.}(2005){Hjorth} {et~al.}}]{Hjorth+05}
{Hjorth} J., {et~al.}, 2005, \nat, 437, 859

\bibitem[{{Janka} {et~al}\mbox{.}(1999){Janka}, {Eberl}, {Ruffert}, \&
  {Fryer}}]{Janka+99}
{Janka} H., {Eberl} T., {Ruffert} M., {Fryer} C.~L., 1999, \apjl, 527, L39

\bibitem[{{Jun}(1998)}]{jun98}
{Jun} B.-I., 1998, \apj, 499, 282

\bibitem[{{Kann} {et~al}\mbox{.}(2008){Kann} {et~al.}}]{Kann+08}
{Kann} D.~A., {et~al.}, 2008, ArXiv e-prints

\bibitem[{{Kasen} \& {Bildsten}(2010)}]{Kasen&Bildsten10}
{Kasen} D., {Bildsten} L., 2010, \apj, 717, 245

\bibitem[{{Kasliwal} {et~al}\mbox{.}(2010){Kasliwal}, {Cenko}, {Kulkarni},
  {Ofek}, {Quimby}, {Rau}, {Caltech}, {Berkeley}, \& {Garching}}]{Kasliwal+10}
{Kasliwal} M.~M. {et~al.}, 2010, ArXiv e-prints

\bibitem[{{Kirk} \& {Skj{\ae}raasen}(2003)}]{Kirk&Skjaerasen03}
{Kirk} J.~G., {Skj{\ae}raasen} O., 2003, \apj, 591, 366

\bibitem[{{Komissarov}(2010)}]{komissarov10}
{Komissarov} S.~S., 2010, ArXiv e-prints

\bibitem[{{Komissarov} \& {Barkov}(2007)}]{Komissarov&Barkov07}
{Komissarov} S.~S., {Barkov} M.~V., 2007, \mnras, 382, 1029

\bibitem[{{K{\"o}nigl} \& {Granot}(2002)}]{Konigl&Granot02}
{K{\"o}nigl} A., {Granot} J., 2002, \apj, 574, 134

\bibitem[{{Kouveliotou} {et~al}\mbox{.}(1993){Kouveliotou}, {Meegan},
  {Fishman}, {Bhat}, {Briggs}, {Koshut}, {Paciesas}, \&
  {Pendleton}}]{kouveliotou93}
{Kouveliotou} C., {Meegan} C.~A., {Fishman} G.~J., {Bhat} N.~P., {Briggs}
  M.~S., {Koshut} T.~M., {Paciesas} W.~S., {Pendleton} G.~N., 1993, \apjl, 413,
  L101

\bibitem[{{Lazzati}, {Morsony} \& {Begelman}(2010){Lazzati}, {Morsony}, \&
  {Begelman}}]{lazzati10}
{Lazzati} D., {Morsony} B.~J., {Begelman} M.~C., 2010, \apj, 717, 239

\bibitem[{{Lee} \& {Ramirez-Ruiz}(2007)}]{lee07}
{Lee} W.~H., {Ramirez-Ruiz} E., 2007, New Journal of Physics, 9, 17

\bibitem[{{Lee}, {Ramirez-Ruiz} \& {L{\'o}pez-C{\'a}mara}(2009){Lee},
  {Ramirez-Ruiz}, \& {L{\'o}pez-C{\'a}mara}}]{Lee+09}
{Lee} W.~H., {Ramirez-Ruiz} E., {L{\'o}pez-C{\'a}mara} D., 2009, \apjl, 699,
  L93

\bibitem[{{Leibler} \& {Berger}(2010)}]{Leibler&Berger10}
{Leibler} C.~N., {Berger} E., 2010, \apj, 725, 1202

\bibitem[{{Levesque} {et~al}\mbox{.}(2010){Levesque}, {Kewley}, {Berger}, \&
  {Jabran Zahid}}]{Levesque+10}
{Levesque} E.~M., {Kewley} L.~J., {Berger} E., {Jabran Zahid} H., 2010, \aj,
  140, 1557

\bibitem[{{Lyons} {et~al}\mbox{.}(2010){Lyons}, {O'Brien}, {Zhang},
  {Willingale}, {Troja}, \& {Starling}}]{Lyons+10}
{Lyons} N., {O'Brien} P.~T., {Zhang} B., {Willingale} R., {Troja} E.,
  {Starling} R.~L.~C., 2010, \mnras, 402, 705

\bibitem[{{MacFadyen}, {Ramirez-Ruiz} \& {Zhang}(2005){MacFadyen},
  {Ramirez-Ruiz}, \& {Zhang}}]{macfadyen+05}
{MacFadyen} A.~I., {Ramirez-Ruiz} E., {Zhang} W., 2005, in Bulletin of the
  American Astronomical Society, Vol.~37, American Astronomical Society Meeting
  Abstracts, pp. 151.04--+

\bibitem[{{MacFadyen} \& {Woosley}(1999)}]{macfadyen99}
{MacFadyen} A.~I., {Woosley} S.~E., 1999, \apj, 524, 262

\bibitem[{{Marietta}, {Burrows} \& {Fryxell}(2000){Marietta}, {Burrows}, \&
  {Fryxell}}]{marietta+00}
{Marietta} E., {Burrows} A., {Fryxell} B., 2000, \apjs, 128, 615

\bibitem[{{Matzner}(2003)}]{matzner03}
{Matzner} C.~D., 2003, \mnras, 345, 575

\bibitem[{{Metzger} {et~al}\mbox{.}(2010{\natexlab{a}}){Metzger}, {Arcones},
  {Quataert}, \& {Mart{\'{\i}}nez-Pinedo}}]{Metzger+10a}
{Metzger} B.~D., {Arcones} A., {Quataert} E., {Mart{\'{\i}}nez-Pinedo} G.,
  2010{\natexlab{a}}, \mnras, 402, 2771

\bibitem[{{Metzger} {et~al}\mbox{.}(2011){Metzger}, {Giannios}, {Thompson},
  {Bucciantini}, \& {Quataert}}]{metzger10}
{Metzger} B.~D., {Giannios} D., {Thompson} T.~A., {Bucciantini} N., {Quataert}
  E., 2011, \mnras, 413, 2031

\bibitem[{{Metzger} {et~al}\mbox{.}(2010{\natexlab{b}}){Metzger},
  {Mart{\'{\i}}nez-Pinedo}, {Darbha}, {Quataert}, {Arcones}, {Kasen}, {Thomas},
  {Nugent}, {Panov}, \& {Zinner}}]{Metzger+10b}
{Metzger} B.~D. {et~al.}, 2010{\natexlab{b}}, \mnras, 406, 2650

\bibitem[{{Metzger}, {Piro} \& {Quataert}(2008){Metzger}, {Piro}, \&
  {Quataert}}]{Metzger+08b}
{Metzger} B.~D., {Piro} A.~L., {Quataert} E., 2008, \mnras, 390, 781

\bibitem[{{Metzger}, {Piro} \& {Quataert}(2009{\natexlab{a}}){Metzger}, {Piro},
  \& {Quataert}}]{Metzger+09a}
---, 2009{\natexlab{a}}, \mnras, 396, 304

\bibitem[{{Metzger}, {Piro} \& {Quataert}(2009{\natexlab{b}}){Metzger}, {Piro},
  \& {Quataert}}]{Metzger+09b}
---, 2009{\natexlab{b}}, \mnras, 396, 1659

\bibitem[{{Metzger}, {Quataert} \& {Thompson}(2008){Metzger}, {Quataert}, \&
  {Thompson}}]{metzger08}
{Metzger} B.~D., {Quataert} E., {Thompson} T.~A., 2008, \mnras, 385, 1455

\bibitem[{{Metzger}, {Thompson} \& {Quataert}(2007){Metzger}, {Thompson}, \&
  {Quataert}}]{Metzger+07}
{Metzger} B.~D., {Thompson} T.~A., {Quataert} E., 2007, \apj, 659, 561

\bibitem[{{Metzger}, {Thompson} \& {Quataert}(2008){Metzger}, {Thompson}, \&
  {Quataert}}]{Metzger+08a}
---, 2008, \apj, 676, 1130

\bibitem[{{Mizuno} {et~al}\mbox{.}(2011){Mizuno}, {Lyubarsky}, {Nishikawa}, \&
  {Hardee}}]{mizuno11}
{Mizuno} Y., {Lyubarsky} Y., {Nishikawa} K., {Hardee} P.~E., 2011, \apj, 728,
  90

\bibitem[{{Morsony}, {Lazzati} \& {Begelman}(2010){Morsony}, {Lazzati}, \&
  {Begelman}}]{Morsony+10}
{Morsony} B.~J., {Lazzati} D., {Begelman} M.~C., 2010, \apj, 723, 267

\bibitem[{{Nakar}(2007)}]{Nakar07}
{Nakar} E., 2007, PhysRep, 442, 166

\bibitem[{{Narayan}, {Piran} \& {Kumar}(2001){Narayan}, {Piran}, \&
  {Kumar}}]{Narayan+01}
{Narayan} R., {Piran} T., {Kumar} P., 2001, \apj, 557, 949

\bibitem[{{Norris} \& {Bonnell}(2006)}]{Norris&Bonnell06}
{Norris} J.~P., {Bonnell} J.~T., 2006, \apj, 643, 266

\bibitem[{{Norris} \& {Gehrels}(2008)}]{norris08}
{Norris} J.~P., {Gehrels} N., 2008, in American Institute of Physics Conference
  Series, Vol. 1000, American Institute of Physics Conference Series,
  {M.~Galassi, D.~Palmer, \& E.~Fenimore}, ed., pp. 280--283

\bibitem[{{Norris}, {Gehrels} \& {Scargle}(2010){Norris}, {Gehrels}, \&
  {Scargle}}]{Norris+10}
{Norris} J.~P., {Gehrels} N., {Scargle} J.~D., 2010, \apj, 717, 411

\bibitem[{{Norris}, {Gehrels} \& {Scargle}(2011){Norris}, {Gehrels}, \&
  {Scargle}}]{norris11}
---, 2011, ArXiv e-prints

\bibitem[{{Ofek} {et~al}\mbox{.}(2007){Ofek}, {Cenko}, {Gal-Yam}, {Fox},
  {Nakar}, {Rau}, {Frail}, {Kulkarni}, {Price}, {Schmidt}, {Soderberg},
  {Peterson}, {Berger}, {Sharon}, {Shemmer}, {Penprase}, {Chevalier}, {Brown},
  {Burrows}, {Gehrels}, {Harrison}, {Holland}, {Mangano}, {McCarthy}, {Moon},
  {Nousek}, {Persson}, {Piran}, \& {Sari}}]{Ofek+07}
{Ofek} E.~O. {et~al.}, 2007, \apj, 662, 1129

\bibitem[{{{\"O}zel} {et~al}\mbox{.}(2010){{\"O}zel}, {Psaltis}, {Ransom},
  {Demorest}, \& {Alford}}]{Ozel+10}
{{\"O}zel} F., {Psaltis} D., {Ransom} S., {Demorest} P., {Alford} M., 2010,
  \apjl, 724, L199

\bibitem[{{Paczynski}(1986)}]{Paczynski86}
{Paczynski} B., 1986, \apjl, 308, L43

\bibitem[{{Perley} {et~al}\mbox{.}(2009){Perley}, {Metzger}, {Granot},
  {Butler}, {Sakamoto}, {Ramirez-Ruiz}, {Levan}, {Bloom}, {Miller}, {Bunker},
  {Chen}, {Filippenko}, {Gehrels}, {Glazebrook}, {Hall}, {Hurley}, {Kocevski},
  {Li}, {Lopez}, {Norris}, {Piro}, {Poznanski}, {Prochaska}, {Quataert}, \&
  {Tanvir}}]{perley09}
{Perley} D.~A. {et~al.}, 2009, \apj, 696, 1871

\bibitem[{{Piro} \& {Ott}(2011)}]{Piro&Ott11}
{Piro} A.~L., {Ott} C.~D., 2011, ArXiv e-prints

\bibitem[{{Price} \& {Rosswog}(2006)}]{Price&Rosswog06}
{Price} D.~J., {Rosswog} S., 2006, Science, 312, 719

\bibitem[{{Prochaska} {et~al}\mbox{.}(2006){Prochaska} {et~al.}}]{Prochaska+06}
{Prochaska} J.~X., {et~al.}, 2006, \apj, 642, 989

\bibitem[{{Rezzolla} {et~al}\mbox{.}(2011){Rezzolla}, {Giacomazzo}, {Baiotti},
  {Granot}, {Kouveliotou}, \& {Aloy}}]{Rezzolla+11}
{Rezzolla} L., {Giacomazzo} B., {Baiotti} L., {Granot} J., {Kouveliotou} C.,
  {Aloy} M.~A., 2011, \apjl, 732, L6+

\bibitem[{{Rossi} \& {Begelman}(2009)}]{Rossi&Begelman09}
{Rossi} E.~M., {Begelman} M.~C., 2009, \mnras, 392, 1451

\bibitem[{{Rosswog}(2007)}]{Rosswog07}
{Rosswog} S., 2007, \mnras, 376, L48

\bibitem[{{Rowlinson} {et~al}\mbox{.}(2010){Rowlinson}, {O'Brien}, {Tanvir},
  {Zhang}, {Evans}, {Lyons}, {Levan}, {Willingale}, {Page}, {Onal}, {Burrows},
  {Beardmore}, {Ukwatta}, {Berger}, {Hjorth}, {Fruchter}, {Tunnicliffe}, {Fox},
  \& {Cucchiara}}]{rowlinson10}
{Rowlinson} A. {et~al.}, 2010, \mnras, 409, 531

\bibitem[{{Shibata} \& {Taniguchi}(2006)}]{Shibata&Taniguchi06}
{Shibata} M., {Taniguchi} K., 2006, Phys.~Rev.~D, 73, 064027

\bibitem[{{Sironi} \& {Spitkovsky}(2011)}]{Sironi&Spitkovsky11}
{Sironi} L., {Spitkovsky} A., 2011, ArXiv e-prints

\bibitem[{{Soderberg} {et~al}\mbox{.}(2005){Soderberg}, {Kulkarni}, {Fox},
  {Berger}, {Price}, {Cenko}, {Howell}, {Gal-Yam}, {Leonard}, {Frail}, {Moon},
  {Chevalier}, {Hamuy}, {Hurley}, {Kelson}, {Koviak}, {Krzeminski}, {Kumar},
  {MacFadyen}, {McCarthy}, {Park}, {Peterson}, {Phillips}, {Rauch}, {Roth},
  {Schmidt}, \& {Shectman}}]{Soderberg+05}
{Soderberg} A.~M. {et~al.}, 2005, \apj, 627, 877

\bibitem[{{Starling} {et~al}\mbox{.}(2010){Starling}, {Wiersema}, {Levan},
  {Sakamoto}, {Bersier}, {Goldoni}, {Oates}, {Rowlinson}, {Campana},
  {Sollerman}, {Tanvir}, {Malesani}, {Fynbo}, {Covino}, {D'Avanzo}, {O'Brien},
  {Page}, {Osborne}, {Vergani}, {Barthelmy}, {Burrows}, {Cano}, {Curran}, {De
  Pasquale}, {D'Elia}, {Evans}, {Flores}, {Fruchter}, {Garnavich}, {Gehrels},
  {Gorosabel}, {Hjorth}, {Holland}, {van der Horst}, {Hurkett}, {Jakobsson},
  {Kamble}, {Kouveliotou}, {Kuin}, {Kaper}, {Mazzali}, {Nugent}, {Pian},
  {Stamatikos}, {Thoene}, \& {Woosley}}]{starling10}
{Starling} R.~L.~C. {et~al.}, 2010, ArXiv e-prints

\bibitem[{{Tagliaferri} {et~al}\mbox{.}(2005){Tagliaferri}
  {et~al.}}]{Tagliaferri+05}
{Tagliaferri} G., {et~al.}, 2005, \nat, 436, 985

\bibitem[{{Thompson}(2010)}]{Thompson10}
{Thompson} T.~A., 2010, ArXiv e-prints

\bibitem[{{Thompson}, {Chang} \& {Quataert}(2004){Thompson}, {Chang}, \&
  {Quataert}}]{Thompson+04}
{Thompson} T.~A., {Chang} P., {Quataert} E., 2004, \apj, 611, 380

\bibitem[{{Thompson}, {Quataert} \& {Burrows}(2005){Thompson}, {Quataert}, \&
  {Burrows}}]{Thompson+05}
{Thompson} T.~A., {Quataert} E., {Burrows} A., 2005, \apj, 620, 861

\bibitem[{{Troja} {et~al}\mbox{.}(2008){Troja}, {King}, {O'Brien}, {Lyons}, \&
  {Cusumano}}]{Troja+08}
{Troja} E., {King} A.~R., {O'Brien} P.~T., {Lyons} N., {Cusumano} G., 2008,
  \mnras, 385, L10

\bibitem[{{Usov}(1992)}]{Usov92}
{Usov} V.~V., 1992, \nat, 357, 472

\bibitem[{{Uzdensky} \& {MacFadyen}(2007)}]{Uzdensky&MacFadyen07}
{Uzdensky} D.~A., {MacFadyen} A.~I., 2007, \apj, 669, 546

\bibitem[{{Virgili} {et~al}\mbox{.}(2011){Virgili}, {Zhang}, {O'Brien}, \&
  {Troja}}]{virgili11}
{Virgili} F.~J., {Zhang} B., {O'Brien} P., {Troja} E., 2011, \apj, 727, 109

\bibitem[{{Wheeler} {et~al}\mbox{.}(2000){Wheeler}, {Yi}, {H{\"o}flich}, \&
  {Wang}}]{wheeler00}
{Wheeler} J.~C., {Yi} I., {H{\"o}flich} P., {Wang} L., 2000, \apj, 537, 810

\bibitem[{{Woosley}(2010)}]{Woosley10}
{Woosley} S.~E., 2010, \apjl, 719, L204

\bibitem[{{Woosley} \& {Baron}(1992)}]{Woosley&Baron92}
{Woosley} S.~E., {Baron} E., 1992, \apj, 391, 228

\bibitem[{{Zhang}(2007)}]{zhang07}
{Zhang} B., 2007, Chin. J. of A\&A, 7, 1

\bibitem[{{Zhang} {et~al}\mbox{.}(2007){Zhang}, {Zhang}, {Liang}, {Gehrels},
  {Burrows}, \& {M{\'e}sz{\'a}ros}}]{Zhang+07}
{Zhang} B., {Zhang} B.-B., {Liang} E.-W., {Gehrels} N., {Burrows} D.~N.,
  {M{\'e}sz{\'a}ros} P., 2007, \apjl, 655, L25

\end{thebibliography}
\bibliographystyle{mn2e}

\label{lastpage}

\end{document}